\documentclass[sn-mathphys-num]{sn-jnl}

\usepackage{graphicx}%
\usepackage{multirow}%
\usepackage{amsmath,amssymb,amsfonts}%
\usepackage{amsthm}%
\usepackage[title]{appendix}%
\usepackage{xcolor}%
\usepackage{textcomp}%
\usepackage{manyfoot}%
\usepackage{booktabs}%
\usepackage{algorithm}%
\usepackage{algorithmicx}%
\usepackage{algpseudocode}%
\usepackage{listings}%

\graphicspath{{figs/}}

\begin{document}

\title[Ethical-Lens (E-LENS)]{E-LENS: User Requirements-Oriented AI Ethics Assurance}


\author*{\fnm{Jianlong} \sur{Zhou}}\email{jianlong.zhou@uts.edu.au}

\author{\fnm{Fang} \sur{Chen}}\email{fang.chen@uts.edu.au}

\affil{\orgdiv{Human-Centered AI Lab, Data Science Institute}, \orgname{University of Technology Sydney}, \orgaddress{\city{Sydney}, \postcode{2007}, \state{NSW}, \country{Australia}}}


\abstract{Despite the much proliferation of AI ethical principles in recent years, there is a challenge of assuring AI ethics with current AI ethics frameworks in real-world applications. While system safety has emerged as a distinct discipline for a long time, originated from safety concerns in early aircraft manufacturing. The safety assurance is now an indispensable component in safety critical domains. Motivated by the assurance approaches for safety-critical systems such as aviation, this paper introduces the concept of AI ethics assurance cases into the AI ethics assurance. Three pillars of user requirements, evidence, and validation are proposed as key components and integrated into AI ethics assurance cases for a new approach of user requirements-oriented AI ethics assurance. The user requirements-oriented AI ethics assurance case is set up based on three pillars and hazard analysis methods used in the safety assurance of safety-critical systems. This paper also proposes a platform named Ethical-Lens (E-LENS) to implement the user requirements-oriented AI ethics assurance approach. The proposed user requirements-based E-LENS platform is then applied to assure AI ethics of an AI-driven human resource shortlisting system as a case study to show the effectiveness of the proposed approach.}




\maketitle

\section{Introduction} 

Artificial Intelligence (AI) is considered as a computer system with at least one of capabilities of perception (e.g. tactile, textual, visual, and audio), decision-making (e.g. financial investment), prediction (e.g. crop yield and health risk), pattern recognition (e.g. anomaly identification), interactive communication (e.g. chat bots or social robots), and many other intelligent functions \cite{vinuesa2020role,zhou2023ai}. These impressive capabilities allow AI perform tasks intelligently comparable with the human intelligence and even much better than humans, which motivates wide applications of AI in various domains such as agriculture, public services, healthcare, education, and others to enhance management performance and promote productivity. AI has been becoming an indispensable component in people's everyday life. However, AI systems are fast, complex, different from humans, and may change via learning, resulting in the societal impacts at a large scale. Furthermore, encased in a black box, contemporary AI solutions often obscure their inner workings in a way. And, ultimately, it is difficult to have trust and confidence in something we do not truly, genuinely, understand. Their behaviour can also be difficult to monitor, validate, predict, and explain. Moreover, AI needs incredible volumes of complex interconnected data (including personal data) for model building, while biases, errors and bad habits in data are being absorbed into most AI systems. The use of personal data also puts people at risk. As a result, AI faces significant ethical challenges ranging from data privacy, explainability, ownership, and privacy, trust, fairness, to accountability and others. In short, two core questions need to be answered on these ethical concerns: 

\begin{itemize}
    \item What is the right thing to do by AI?
    \item How do we know for whom and whether AI is ``good''?
\end{itemize}

In order to mitigate ethical concerns of AI, various AI ethical frameworks and guidelines have been proposed by national and international organisations as well as non-profit organisations and professional associations \cite{zhou2020survey}. For example, the AI ethical principles identified by Data61 in Australia include: human-centred values; human, social and environmental wellbeing; fairness; privacy protection and security; transparency and explainability; reliability and safety; contestability; and accountability \cite{dawson2019artificial}. IEEE's AI ethical principles include: wellbeing; human rights; transparency; accountability; data agency; effectiveness; awareness of misuse; and competence \cite{chatila2019ieee}. It was found that no single ethical principle is explicitly agreed by all published ethical guidelines. However, it shows a convergence around the principles of beneficence, responsibility, transparency, justice and fairness, non-maleficence, privacy, trust, sustainability, dignity, freedom and autonomy, and solidarity in the global policy landscape \cite{jobin2019global}. Furthermore, organisations identified mandatory AI ethical principles based on those convergent ethical principles: community benefit, transparency, fairness, privacy and security, and accountability \cite{digitalnsw2024}.

Despite the much proliferation of AI ethical principles in recent years, there is a challenge of assuring AI ethics with current AI ethics frameworks in real-world applications \cite{zhou2023ai}. The current frameworks of AI ethics are usually abstract and provide limited approaches for developers and researchers of algorithms to apply them in practical applications. For example, it was found that 75\% of ethical frameworks only contain general principles with very little details, and more than 80\% of them offer no or very little information on practical applications \cite{belisle2022artificial,zicari2022assess}.
Checklist-style questionnaire is one of widely accepted tools for the operationalisation of AI ethical principles.
The Assessment List for Trustworthy Artificial Intelligence (ALTAI) published by the High-Level Expert Group on Artificial Intelligence (HLEG-AI) of European Commission in 2020 is one of checklist-style questionnaire examples \cite{ala2020assessment}. Han and Choi \cite{han2022checklist} devise a checklist to review trustworthiness of AI based on ethical principles of explainability, fairness, robustness, Safety, and transparency. However, the checklist questions are mostly simple Yes/No validations and there are no further detailed insights/suggestions, which are not enough for computational models. 
The documentation is another intervention that has been adapted and developed to communicate key facts about machine learning (ML) systems. Datasets \cite{gebru2021datasheets}, FactSheets \cite{arnold2019factsheets}, and Method Cards \cite{adkins2022method} are documentation tools proposed to provide key details about datasets, models and other details about AI development.
Furthermore, frameworks are developed by providing an integrated process to assure AI ethics. For example, Z-Inspection \cite{zicari2021z} outlines a process with three phases of setup, assess, and resolve phases to assess trustworthy AI.
However, these endeavours also often only focus on general questions and do not consider requirements of specific domains. They pay little attention to user requirements on AI solutions, which is the key for both AI solutions and AI ethics.

Moreover, system safety has emerged as a distinct discipline for a long time, originated from safety concerns in early aircraft manufacturing \cite{rinehart2015current}. The safety assurance is now an indispensable component in safety critical domains such as nuclear, aviation, energy, rail, finance \cite{rinehart2015current}, which is characterised by high levels of safety criticality, and technical and operational complexity.
For example, Sharifi et al. \cite{sharifi2022towards} use requirements engineering techniques and standards, which are common in safety-critical sectors, to set up a set of requirements-based guidelines to build a rigorous certification for FinTech systems. Goal modeling, systemic safety assessment, and assurance case development are used in these guidelines. Safety cases try to provide an established approach for validating the safety of systems through an explicit argument supported by compelling evidence \cite{picardi2020assurance}.

Motivated by the assurance approaches for safety-critical systems such as aviation \cite{rinehart2015current}, this paper introduces the concept of assurance cases into AI ethics assurance. Three pillars of user requirements, evidence, and validation (REV) are proposed and integrated into AI ethics assurance cases for a new approach of user requirements-oriented AI ethics assurance. User requirements on AI ethics are user's specific interest in ethical concerns and they are the basis for the assurance. The clear identification of user requirements on ethics allows efficient assurance of targeted ethical issues instead of non-focused general ethical problems. Furthermore, the AI ethics in real-world applications must be assured with enough evidence that AI models are used ethically. Such evidence must be sufficient to explain why the AI system can be trusted ethically for its applications. The evidence is then validated and verified by experts in AI ethics and/or regulators so that AI systems can be endorsed and used ethically. Based on these three pillars, the AI ethics assurance case is set up using System Theoretic Process Analysis (STPA) \cite{leveson2016engineering} which is a hazard analysis method widely used in the safety assurance of safety-critical systems. This paper also proposes a platform named Ethical-Lens (E-LENS) to implement the user requirements-oriented AI ethics assurance approach. 


The primary contributions of this paper include:
\begin{itemize}
  \item The concept of assurance case widely used in safety-critical systems is introduced into AI ethics assurance to validate AI ethical principles.
  \item A new approach of user requirements-oriented AI ethics assurance is proposed by integrating three pillars of user requirements, evidence, and validation into AI ethics assurance cases.
  \item A new platform of E-LENS is proposed to implement the user requirements-oriented AI ethics assurance approach. 
\end{itemize}

\section{RELATED WORK}

\subsection{AI ethics operationalisation}

Han and Choi \cite{han2022checklist} formulate the considerations for validating a trustworthy AI and devise a checklist to review trustworthiness of AI. The checklist items are based on ethical principles of explainability, fairness, robustness, Safety, and transparency throughout the AI life cycle. The High-Level Expert Group on Artificial Intelligence (HLEG-AI) of European Commission published the Assessment List for Trustworthy Artificial Intelligence (ALTAI) in 2020, which represents a significant step in terms of AI ethics assurance. The ALTAI checklist tool enables users to conduct assessments of AI systems about the governance to the ethical principles \cite{ala2020assessment}. Some studies have used ALTAI in examining specific contexts, such as how users should achieve ethical AI, or how the ALTAI list can be applied in driver-assistance systems \cite{borg2021exploring,gardner2022ethical}. Radclyffe et al. \cite{radclyffe2023assessment} review the pros and cons of the ALTAI checklist tool. The major strength lies in its capability to implement the HLEG-AI guideline into an objective tool that can measure the assurance results. The weakness shows that the ALTAI checklist tool does not provide guidance for an organization's level of development on AI ethics. ALTAI also lacks actionable recommendations, particularly aimed at industry developments and practices on AI technologies \cite{radclyffe2023assessment}. It also lacks quantitative evaluations of ethics on some ethical principles such as explainability and fairness. Furthermore, the checklist questions are mostly only simple Yes/No validations and there are no further detailed insights/suggestions, which are not enough for AI computational models. 

The documentation is a practical tool to provide clarifications for stakeholders in software engineering. Such documentation techniques haven been adapted and developed to communicate key facts about machine learning systems. For example, FactSheets \cite{arnold2019factsheets} provide declarations of conformity for AI solutions with multi-dimensional details that capture various aspects of the AI model and its development processes to make it trustworthy by customers. Datasheets for Datasets \cite{gebru2021datasheets} aims to provide key information about the datasets used to develop machine learning models. Model Cards \cite{mitchell2019model} are short documents that provide benchmarked evaluations in a variety of conditions for trained machine learning models. Model Cards also provide details of the model performance evaluation procedures, the context where models are to be used, and other relevant information. While the information provided by above methods is essential for stakeholders to evaluate whether an AI system meets their requirements, they are still less actionable. For example, machine learning developers need guidance to mitigate potential shortcomings to improve the system’s performance while above methods may not provide such guidance. Method Cards \cite{adkins2022method} aim to communicate key information with both prescriptive and descriptive elements about machine learning methods to guide machine learning developers throughout the process of model development to ensure that they are able to use those methods properly. Despite the usefulness of these approaches toward transparency and accountability of the machine learning system, they do not cover all aspects needed for the assurance of ethics and are limited from a user requirements' perspective on data and machine learning models.

Frameworks are also developed for the assurance of AI ethics. For example, Zicari et al. \cite{zicari2021z} outline a process named Z-Inspection to assess trustworthy AI. The process includes three phases of setup, assess, and resolve phases. The setup phase uses a catalog of questions to define an understanding of the expectations of various stakeholders involved in the assessment. The assess phase analyses the usage scenarios,  maps the ethical issues to AI ethical principles, and verifies AI ethical principles. Z-Inspection uses the definition of trustworthy AI given by the high-level European Commission's expert group on AI. The term of trustworthiness is more on the machine learning model's performance \cite{ashmore2021assuring} instead of ethics.


Standards play a key role in AI ethics operationalisation. There are different voluntary standardization and certification initiatives in AI, notably from IEEE \cite{ieee2019draft,ieee2020recommended} and NIST \cite{schwartz2022towards}. Australia published the Voluntary AI Safety Standard in August 2024, which gives practical guidance to Australian organisations on how to safely and responsibly use and innovate with AI \cite{national2024voluntary}. The standards aim to ensure that the development and deployment of AI systems is safe and can be relied on.
Furthermore, organisations and big companies have published different tools from the perspective of design of AI. For example, Microsoft \footnote{https://www.microsoft.com/en-us/haxtoolkit/} published a human-AI experiences (HAX) toolkit with a series of guidelines for the design of AI solutions. Google's People+AI \footnote{https://pair.withgoogle.com/guidebook} proposed design patterns and guidance for human-centred AI across the AI product development flow. Liao et al. \cite{liao2021question} from IBM Research present a question-driven design process for explainable AI user experiences. The question-driven design process considers choices of explainable AI techniques, user needs, design, and evaluation of user experiences in the user questions. These work are important steps for the assurance of AI ethics. However, they still lack actionable tools for the AI ethics assurance or do not consider specific user requirements on ethics for targeted assurance.

\subsection{Assurance case for mission critical systems}

Machine learning is often a major component in cyber-physical systems (CPSs) in safety-critical domains (e.g. financial, medical, and automotive domains). The failure of machine learning driven software in such systems can lead to the loss of human lives and/or spend significant costs to fix problems. The assurance of the safety for these systems is conducted from different perspectives: 
1) Assessment of confidence and uncertainty \cite{duan2017reasoning}. The confidence in an assurance case is defined as ``the quality or state of being certain that the assurance case is appropriately and effectively structured, and correct'' \cite{grigorova2013taking}. One approach to gain confidence in the assurance case is to analyse uncertainty, which can be dealt with through qualitative analysis or quantitative analysis. 
2) Assessment of structure and content: software tools are developed to support assessors for processing assurance case documents such as structural checks and content checks: three categories of methods  (correctness and completeness checks, structural constraints, user queries) are summarised for assessing an assurance case's structure, and five categories (argument assessment, evidence assessment, assessment interaction, assessment tracking, assessment reporting) are summarised for assessing an assurance case's content \cite{maksimov2019survey}. 
3) Security assessment \cite{mohamad2021security}. Security assurance cases are used to evaluate the security properties of a software system. It requires a thorough threat analysis throughout the software structure and at different stages of the software development. 
4) Assurance weakeners \cite{shahandashti2023prisma}: The presence of assurance weakeners such as assurance deficits and logical fallacies in assurance cases indicates incomplete knowledge, evidence, or gaps in verification. These weakeners can affect user confidence in assurance arguments, interfering the successful assurance of the system. Shahandashti et al. \cite{shahandashti2023prisma} conduct a systematic review and report a taxonomy that categorizes assurance weakeners and related management approaches at the modeling level. This taxonomy can help practitioners to have better understanding of weakeners of existing assurance solutions. Based on the understanding, practitioners can devise more effective solutions to manage assurance weakeners.

Machine learning development is an iterative process. Therefore, an assurance case needs to continuously update related assurance artifacts such as datasets information and performance changes. Carlan et al. \cite{carlan2022automating} examine how changes in the ML process affect safety arguments, with a focus on the adequacy of the data used to develop ML components. Yap \cite{yap2021towards} presents a process-based framework for certifying trustworthy ML systems by adopting a simplified version of the common criteria used for certifying information technology and security software. Users can rely on the certification process-based framework to build confidence in the trustworthiness of an ML system. The certification is divided into two components: certifying basic system components which are covered by a disclosure such as FactSheets \cite{arnold2019factsheets} and certifying desired properties.

Picardi et al. \cite{picardi2020assurance} introduce patterns for developing assurance arguments to demonstrate the safety of ML components. These argument patterns offer reusable templates for the necessary claims in an argument.
They also develop a process which comprises of five ML lifecycle phases (requirements elicitation, data management, model learning, model verification, and model deployment) and four-level assurance and development activities (assurance artifacts, assurance activities, development activities, and system artifacts), to establish the argument patterns and construct the assurance case for ML components. Assurance requirements for each ML lifecycle phase are derived based on a set of desiderata for each lifecycle phase \cite{ashmore2021assuring}. The patterns and process serve as a practical guide for the justifiable deployment of ML models in safety-related systems.
These previous work for mission critical systems motivates us to develop AI assurance cases by considering user requirements and evidence in AI ethics. 

\section{PRELIMINARIES}


This section introduces concepts of assurances cases and System Theoretic Process Analysis that are used to set up AI ethics assurance approaches proposed in this paper.

\subsection{Assurance cases}

The term ``assurance'' is defined by the Cambridge Dictionary as ``a promise to tell something to someone confidently or firmly, or a promise to cause someone to feel certain by removing doubt'', and has long been used alongside objectives for safety, security, and other concerns \cite{rinehart2015current}. The term ``assurance case'', a generalization of a safety case, has become familiar to many practitioners \cite{ieee2022standard}. An assurance case (AC) is defined as ``a collection of verifiable claims, arguments, and evidence assembled to confirm that a particular system/service aligns with given requirements'' to solve the question of ``are we sufficiently certain that this system is acceptably [safe, secure, reliable, etc.]?'' \cite{mansourov2010system}. An assurance case is usually represented either as plain text (unstructured natural language text, semi-structured text) or graphical notations that allows communication among various system stakeholders (e.g., suppliers, users, regulators). Many notations and tools have been proposed to create assurance arguments such as Goal Structuring Notation (GSN) \cite{chelouati2023graphical}. 

Furthermore, AI systems are sociotechnical, consisting of significant human and software elements. Goal modeling has proven useful in capturing stakeholder needs and documenting arguments and justifications for compliance, trade-offs, and decision-making in such systems \cite{horkoff2019goal}. 
The User Requirements Notation (URN) \cite{itu2018151,dongmo2023improved}, which is based on goal modeling approach, is one of popular notations that are used in the safety assurance of mission critical systems \cite{sharifi2022towards}. 
URN \cite{itu2018151,dongmo2023improved} is a notation used for eliciting, analyzing, specifying, and validating requirements. It enables software and requirements engineers to identify and define requirements for both proposed and evolving systems, as well as analyze them for correctness and completeness. It combines goal and process views used in requirements engineering activities, regulatory compliance and regulatory intelligence \cite{sharifi2022towards}. It is widely used in telecommunications systems, services and business processes, as well as the high-level design of systems.

\begin{figure}[!htb]
  \centering
  \includegraphics[width=0.99\linewidth]{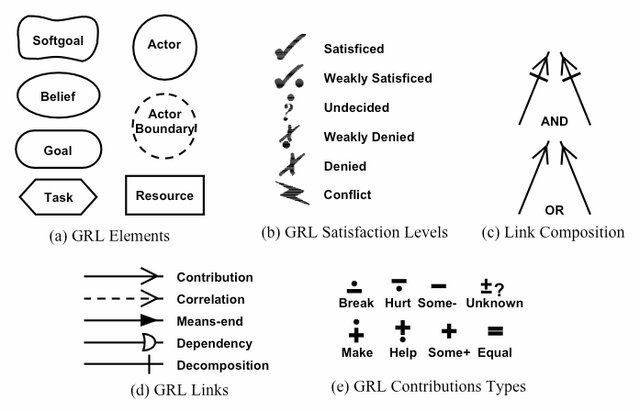}
    \caption{Categories of concepts in Goal-Oriented Requirements Language \cite{weiss2007business}.}
\label{fig:Elements-of-GRL}       
\end{figure}

URN integrates two sub-languages: the Goal-oriented Requirement Language (GRL), for modeling actors and their intentions, and Use Case Maps (UCM), for describing scenarios and architectures. GRL supports goal-oriented modeling and reasoning, particularly for non-functional requirements and quality attributes. It features five main categories of concepts (see Figure~\ref{fig:Elements-of-GRL}): elements, satisfaction levels, link composition, links and contribution types. UCM focuses on visually representing the causal flow of behavior, optionally overlaid on a structure of components. It includes process-oriented concepts such as start and end points, responsibilities or activities, their corresponding system components or actors, and various relationships, including sequencing, guarded choices, concurrency, process decomposition, timers, and more \cite{amyot2011user}. 



\subsection{System Theoretic Process Analysis}

\begin{figure*}[!htb]
  \centering
  \includegraphics[width=0.99\linewidth]{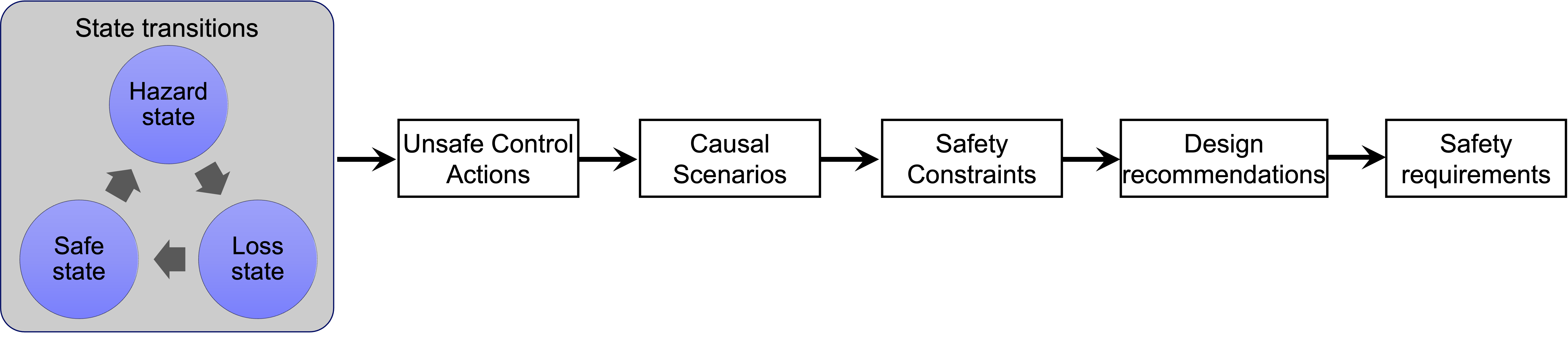}
    \caption{System Theoretic Process Analysis pipeline.}
\label{fig:STPA-process}       
\end{figure*}

System Theoretic Process Analysis (STPA) \cite{leveson2016engineering} is a hazard analysis method based on the System Theoretic Accident Model and Processes (STAMP) that evaluates potential systemic events where undesirable outcomes may occur. It offers a systematic approach to system safety and ensures clear traceability between hazards and design. Figure~\ref{fig:STPA-process} illustrates the STPA pipeline. In this pipeline, STPA defines three system states: 1) loss states, which are unacceptable to stakeholders (e.g., loss of fairness, assets, or reputation); 2) hazard states, where no loss has occurred yet but which could lead to a loss state; and 3) safe states, which are acceptable to stakeholders. STPA uses these concepts within a hierarchical control structure to identify potential pathways through which the system might transition to a hazard state. As shown in Figure~\ref{fig:STPA-process}, the control actions that  lead to hazards referred to as Unsafe Control Actions (UCAs). The UCAs are used as a basis to identify causal scenarios that might cause the UCAs. Safety constraints are derived from causal scenarios, which are then used to define design recommendations to mitigate the hazards \cite{sharifi2022towards}. System safety requirements are identified for the overall system safety assurance.

STPA offers a broad definition of loss that can be adapted to different types of losses in other safety-critical areas. Additionally, STPA provides a methodology for systematically defining system controls that are traceable to the losses stakeholders aim to prevent \cite{sharifi2022towards}. Such features much benefit a requirements-based certification approach.

\section{User Requirements-Oriented AI Ethics Assurance}

This section first introduces the concept of AI ethics assurance case. A user requirements-oriented approach for AI ethics assurance framework is then proposed. The detailed AI ethics assurance process is also presented in this section. The traceable link between user requirements on ethics and AI ethics assurance is the focus of this section.

\subsection{AI Ethics Assurance Case}

Assuring AI ethics requires more than simple reports from validators. A rigorous and systematic approach to capturing and communicating the rationale throughout the AI lifecycle is essential for certifying AI systems.
In this paper, the concept of assurance case is adapted into AI ethics for the AI ethics assurance. An AI ethics assurance case is a comprehensive, defensible, and valid justification of the ethical considerations for a specific AI system and its application. The assurance case therefore provides the grounds for user confidence and trust in AI solutions from the ethical perspective. Similar to system safety assurances, the AI ethics assurance case acts as the basis for the decision making in certification of AI solutions.

Adapted from the safety assurance cases \cite{picardi2020assurance}, the AI ethics assurance claim is that the AI model meets the defined ethical requirements, including ethical principles relevant to the specified operating environment. The claim focuses on the specific AI ethical principles to reflect the AI model's impact on the overall system's ethics. Furthermore, the verification evidence that supports the claim must be specific evidence that shows the AI ethical principles are satisfied.
The assurance argument for an AI model should be viewed as a component of a broader assurance argument addressing the safety of the entire system.
AI ethics assurance involves multiple concerns (business, regulations, AI, data) and must clearly convey compliance and risk assessments to certification authorities.

\subsection{AI Ethics Assurance Framework -- A User Requirements-Oriented Approach}

\begin{figure*}[!htb]
  \centering
  \includegraphics[width=0.99\linewidth]{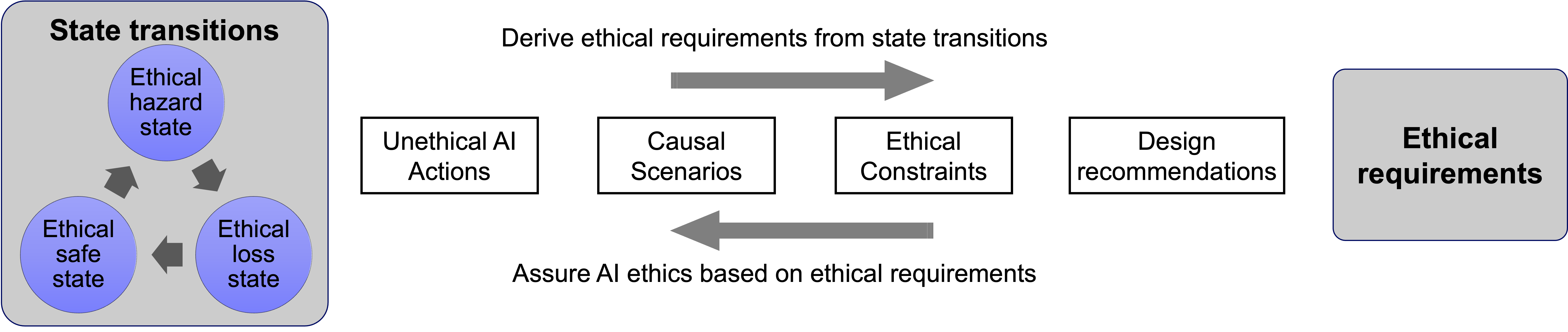}
    \caption{Ethical process analysis from ethical requirements' perspective.}
\label{fig:ethical-STPA}       
\end{figure*}

Motivated by the system theoretic process analysis in safety critical systems \cite{sharifi2022towards}, this paper proposes a requirements-oriented approach for AI ethics assurance cases, focusing on addressing ethical requirements rather than merely arguing that AI solutions are free from ethical hazards or risks. Figure~\ref{fig:ethical-STPA} shows the ethical process analysis from ethical requirements' perspective. In this process, three states of the AI ethical status are defined: 1) ethical safe state or ethical state which is the safe state in ethics; 2) ethical loss state or loss state that is unacceptable to the stakeholders of the AI system from the ethical perspective, e.g. leak of privacy, lack of transparency; and 3) ethical hazard state or hazard state, where no ethical loss has occurred yet but which could transition to an ethical loss state. Similar to STPA for safety assurance cases \cite{sharifi2022towards}, these states can be transitioned from one state to another. A hierarchical control structure is used for such state transitions. This ethical process analysis is divided into two processes (see Figure~\ref{fig:ethical-STPA}): derive design recommendations from state transitions, and assure AI ethics based on ethical requirements. In the process of deriving design recommendations from state transitions, the Unethical AI Actions (UAIAs) that can lead to ethical hazards are firstly identified. The UAIAs are used as a basis to identify ethical causal scenarios that might cause the UAIAs. Ethical constraints are then derived from causal scenarios. Design recommendations are therefore introduced based on constraints. Finally, ethical requirements are identified to mitigate the ethical hazards in the state transitions. On the other side, the ethical requirements are used to assure AI ethics by checking components from ethical requirements to UAIAs in the assurance case (see Figure~\ref{fig:ethical-STPA}).

\begin{figure*}[!htb]
  \centering
  \includegraphics[width=0.5\linewidth]{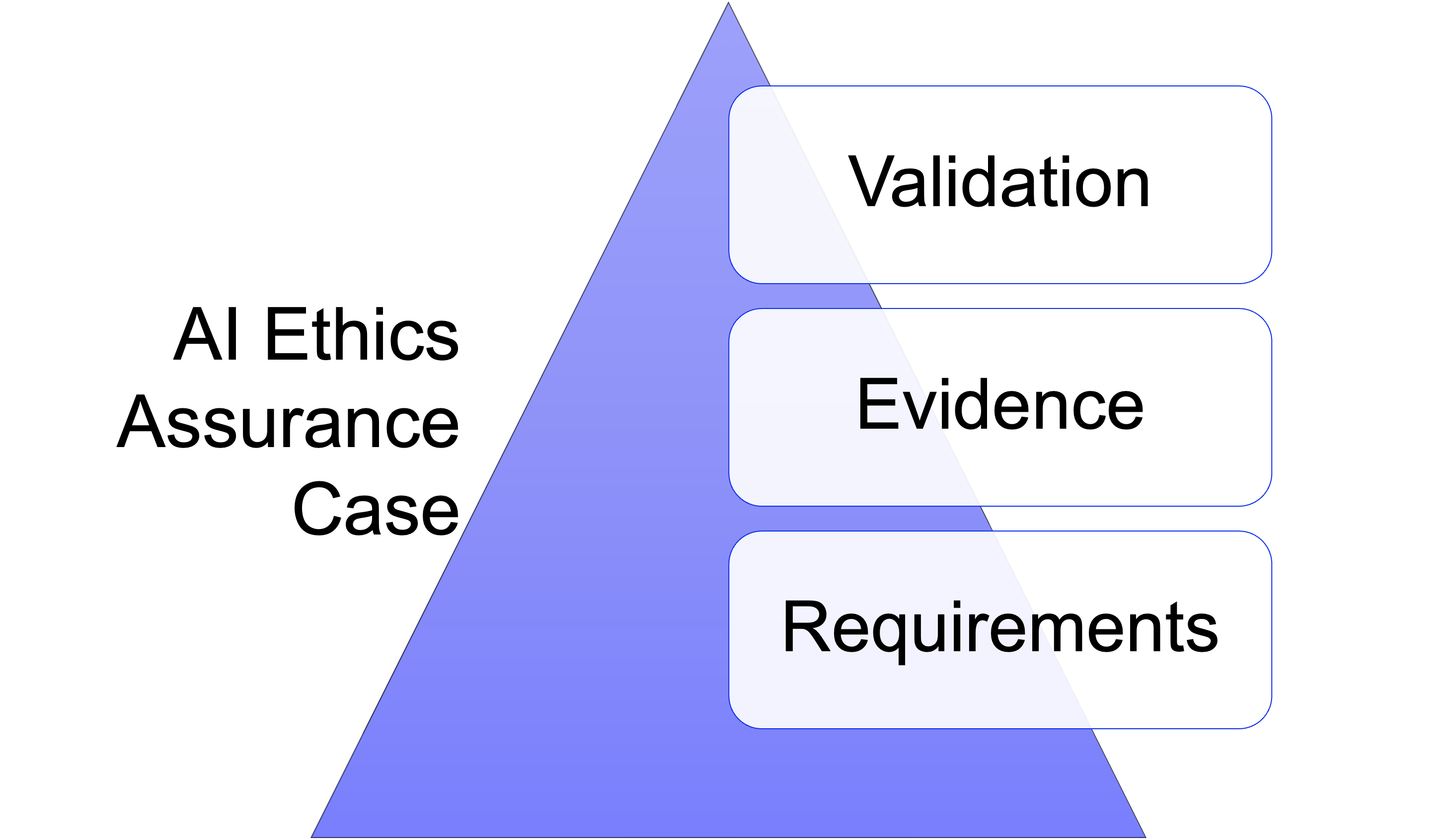}
    \caption{Three pillars in AI ethics assurance cases.}
\label{fig:three-pillars}       
\end{figure*}

Besides requirements, this paper also introduces evidence and validation into the AI ethics assurance framework. These three components of requirements, evidence and validation (REV) are called three pillars of AI ethics assurance cases (see Figure~\ref{fig:three-pillars}). User requirements on AI ethics are user's specific interest in AI ethics and they act as the basis for the assurance. The evidence is any supporting information that is used to justify the ethical trustworthiness of AI systems for their intended applications. The validation is the step that is conducted by experts in AI ethics and/or regulators so that AI systems can be endorsed and used ethically. Based on these three pillars, the AI ethics assurance case is set up using System Theoretic Process Analysis.

Based on the ethical process analysis, URN is used to model the AI ethics assurance process in this paper. This process includes the following steps:


\begin{enumerate}
  \item Identify stakeholders.
  \item Develop a strategic dependency model.
  \item Detail the functional goals.
  \item Conduct STPA.
  \item Create assurance cases.
\end{enumerate}


The details of these steps are described in the following subsections.

\subsection{Identify stakeholders}
The first step in the AI ethics assurance process is to develop a comprehensive understanding of AI solution stakeholders and their dependencies, known as a strategic dependency model in goal-oriented requirements engineering \cite{sharifi2022towards}. Seven groups of stakeholders are identified in the AI ethics assurance cases:
\begin{itemize}
    \item AI suppliers: Those responsible for creating, providing, and supporting the AI system in the assurance.
    \item AI supplier administrators: Those who have administrative responsibility for the AI system.
    \item AI Regulators: Those who are responsible for overseeing and enforcing regulations related to the development, deployment, and use of AI based on the assurance case for the AI system.
    \item System administrators: Those responsible for administrative role on the assurance system (e.g. E-LENS in our study).
    \item Ethics validators: Those who check evidence provided by AI suppliers for the compliance of the AI system on AI ethical principles.
    \item AI users: Those who rely on AI solutions to inform their decisions.
    \item Visitors/General public: The public is given access to the assurance system and the assurance case to understand its function, ensure transparency, and assess any residual risks.
\end{itemize}

\begin{figure*}[!htb]
  \centering
  \includegraphics[width=0.8\linewidth]{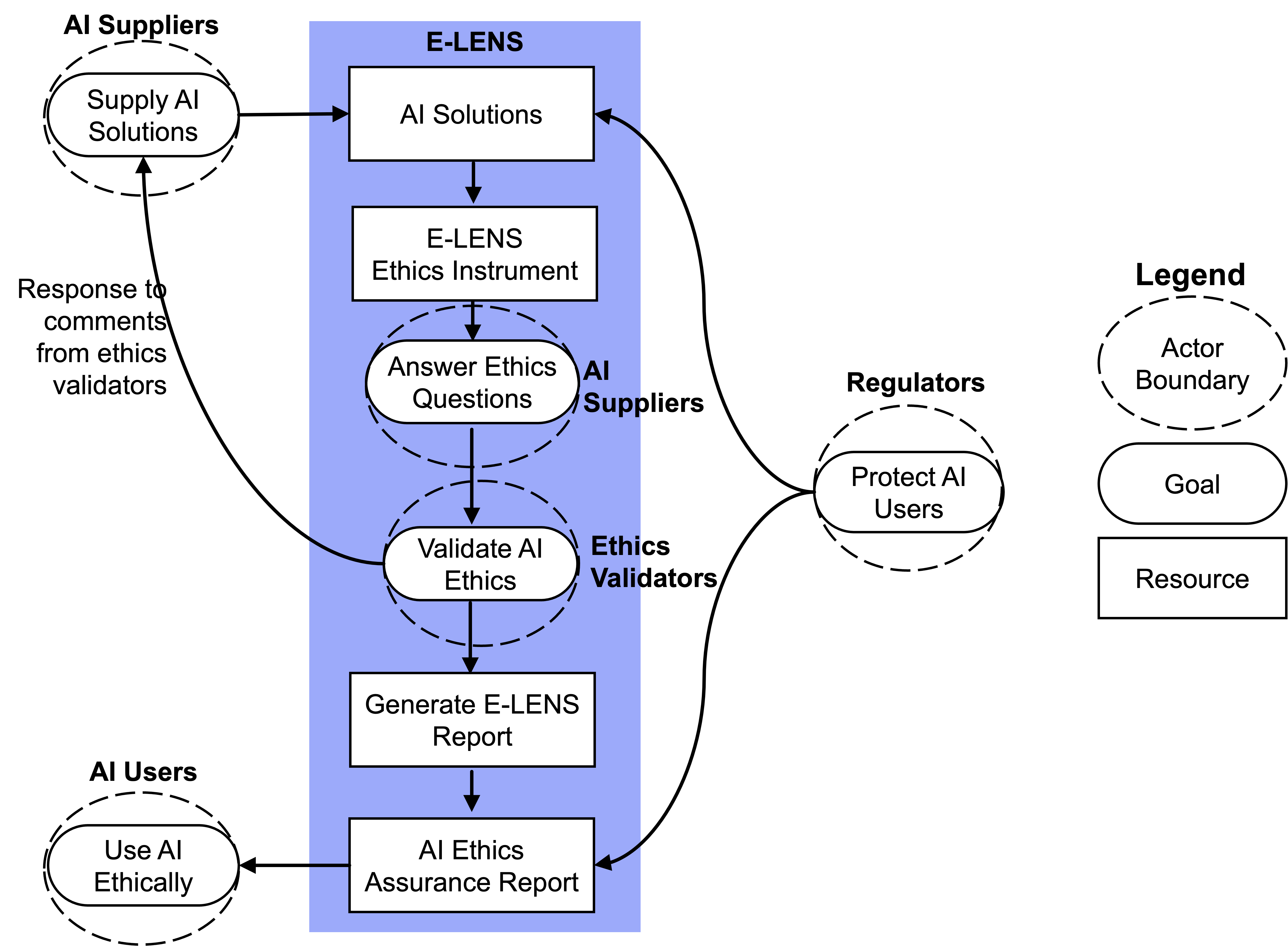}
    \caption{Strategic dependency model in the AI ethics assurance case.}
\label{fig:E-LENS-strategic-denpendency}       
\end{figure*}

Among these stakeholders, AI supplier administrators, system administrators, and visitors/general public are the administrators of assurance systems or general public, and they could not be considered in the strategic dependency model. Figure~\ref{fig:E-LENS-strategic-denpendency} shows the strategic dependency model in the proposed AI ethics assurance case based on the identified stakeholders. It allows for the identification of high-level goals of various stakeholders and their inter-dependencies. 

\subsection{Detail the functional goals}

\begin{figure*}[!htb]
  \centering
  \includegraphics[width=0.99\linewidth]{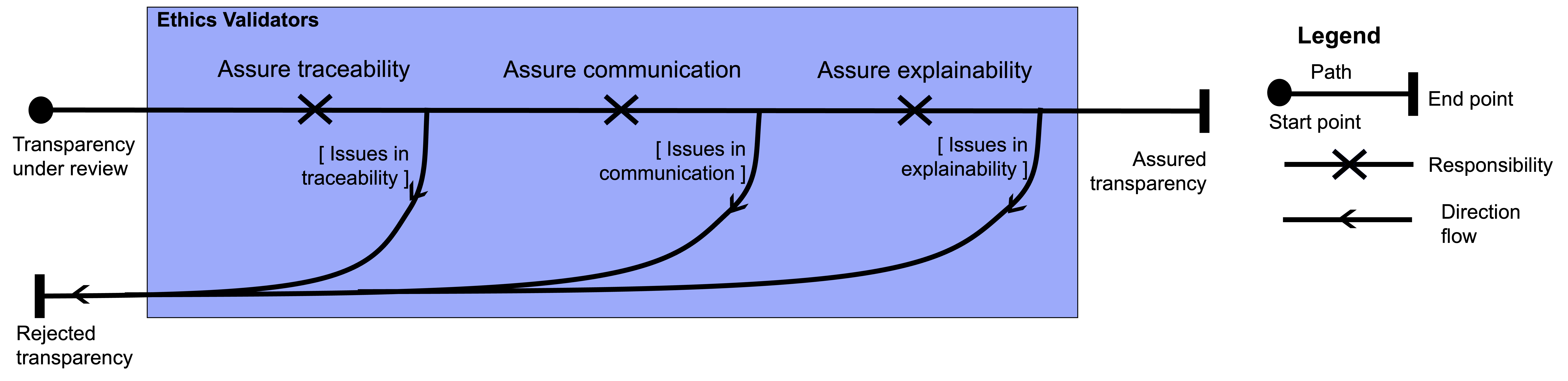}
    \caption{Use Case Map for the transparency.}
\label{fig:user-case-map}       
\end{figure*}

The high-level goals identified in the strategic dependency model can be further refined with the use of UCM described in URN. UCM gives a visual scenario of functional goals of the system and actors' behavior. This section uses the AI ethical principle of transparency as an example to show how URN is used for the AI ethics assurance case. Figure~\ref{fig:user-case-map} shows the UCM for the transparency in the AI ethics assurance case. In this UCM, transparency is assured by checking the traceability, communication and explainability sequentially. The failure of the assurance of any component results in the failure of the assurance of the transparency.

\subsection{Conduct STPA}

With the top-level goals identified in the previous step using UCM, the ethical loss states and their corresponding ethical hazards are defined to establish the ethical boundaries of the AI solution. Table~\ref{tab:list-ethical-losses} shows the list of example ethical losses for transparency, and Table~\ref{tab:list-ethical-hazards} presents the list of example ethical hazards corresponding to ethical losses in transparency in Table~\ref{tab:list-ethical-losses}.

\begin{table*}[!htb]
\caption{List of ethical losses in transparency.}
\begin{tabular*}{31pc}{@{}p{60pt}p{20pt}p{270pt}@{}}
\toprule
Principle segment & ID & Ethical losses \\
\hline
Traceability & L1 &  Loss of documentation related to data, its collection, and preprocessing\\
[3pt] & L2 & Loss of the documentation of the information on the method of training and testing the algorithm\\
[3pt] & L3 & Loss of documentation regarding the data used for testing and validation\\ 
[3pt] & L4 & Loss of the documentation of the outcomes of or decisions taken by the algorithm\\
\hline
Communication & L5 & Loss of clear communication about the AI system's characteristics, limitations, and risks\\
\hline
Explainability & L6 & Loss of the ability to provide an explanation for why the system made a particular choice leading to a specific outcome, in a way that is understandable to all users\\
[3pt] & L7 & Loss of the evaluation of the faithfulness of explanation, i.e. are relevance scores in explanations indicative of ``true" importance\\
[3pt] & L8 & Loss of evaluating the monotonicity of the explanation, meaning whether adding more positive evidence increases the likelihood of classification in the specified class\\
\botrule
\end{tabular*} 
\label{tab:list-ethical-losses}
\end{table*}

\begin{table*}[!htb]
  \caption{List of ethical hazards in transparency.}
  \begin{tabular*}{31pc}{@{}p{60pt}p{20pt}p{235pt}p{20pt}@{}}
  \toprule
  Principle segment & ID & Ethical losses & Links to losses\\
  \hline
  Traceability & H1 &  Required documentation of the information on the data, collection, and preprocessing not obtained & L1\\
  [3pt] & H2 & Required documentation of the information on the method of training and testing the algorithm not obtained & L2\\
  [3pt] & H3 & Required documentation of the information on the data used to test and validate not obtained & L3\\
  [3pt] & H4 & Required documentation of the algorithm's outcomes or decisions was not obtained & L4\\
  \hline
  Communication & H5 & There is no communication regarding the AI system's characteristics, limitations, or risks & L5\\
  \hline
  Explainability & H6 & Explanations on the model or outputs not provided & L6\\
  [3pt] & H7 & Explanations on the model or outputs are wrong & L6\\
  [3pt] & H8 & Evaluation of the faithfulness of explanation not provided & L7\\
  [3pt] & H9 & The faithfulness of explanation is low & L7\\
  [3pt] & H10 & Evaluation of the monotonicity of explanation not provided & L8\\
  [3pt] & H11 & The monotonicity of explanation is low & L8\\
  \botrule
  \end{tabular*} 
  \label{tab:list-ethical-hazards}
  \end{table*}

\begin{figure*}[!htb]
    \centering
    \includegraphics[width=0.99\linewidth]{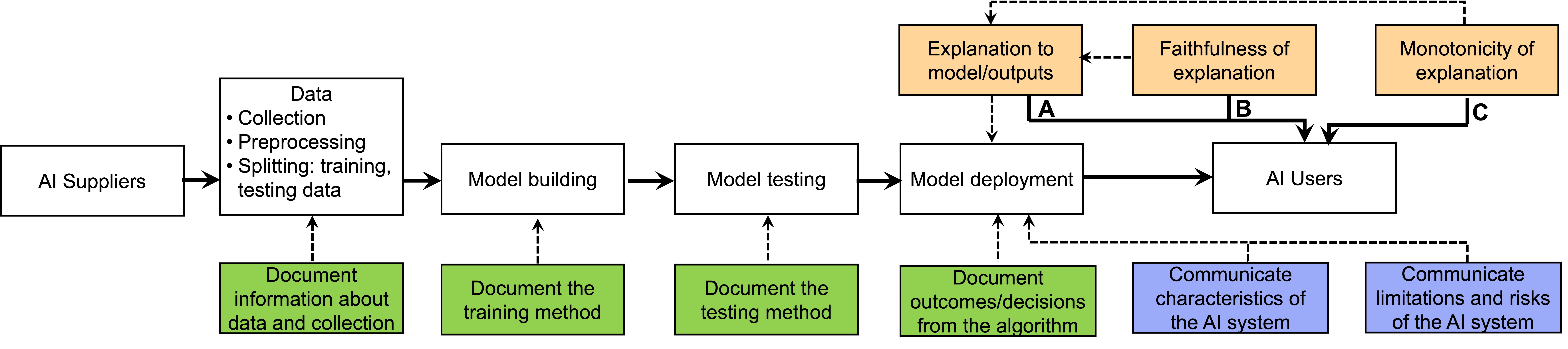}
      \caption{Control diagram regarding transparency for an AI model building in the AI lifecycle.}
  \label{fig:control-diagram-transparency}       
  \end{figure*}

With the definition of ethical losses and hazards, it is necessary to evaluate possible ways in which any ethical losses and hazards can occur. According to the theory of System Theoretic Accident Model (STAM) and Processes used in STPA \cite{sharifi2022towards}, STAM treats systemic qualities as a dynamic control issue. A control diagram would help to visually understand the relations of different components and how they affect each other. Figure~\ref{fig:control-diagram-transparency} shows the control diagram regarding transparency for an AI model building in the AI lifecycle. In this diagram, the green shaded boxes are related to the ethical principle segment of traceability, the purple shaded boxes are related to the ethical principle segment of communication, and the orange shaded boxes are related to the ethical principle segment of explanation. From this control diagram, we can easily see that how different actions affect the transparency of which component in the AI lifecycle, for example, ``Document the training method'' affects the traceability of ``Model building'', ``Explanation to model/outputs'' (Action A) affects the explainability to ``AI users''.

Taking the control action of A (provide ``Explanation to model/outputs'') in Figure~\ref{fig:control-diagram-transparency} as an example, it is necessary to identify unethical AI actions for the action A based on the proposed framework (see Figure~\ref{fig:ethical-STPA}). Table~\ref{tab:list-ethical-UAIA} illustrates examples of UAIAs identified for the actions of A, B, and C (actions can be hazards) when those actions are provided or not provided.

\begin{table*}[!htb]
  \caption{Samples of unethical AI actions (UAIAs) identified for the control actions A, B, and C.}
  \begin{tabular*}{31pc}{@{}p{70pt}p{140pt}p{140pt}@{}}
  \toprule
  Control action & Provided & Not provided \\
  \hline
  A: Explain model/outputs & UAIA101: The AI system provides the wrong explanation. [H7] &  UAIA104: The AI system does not explain the model/outputs. [H6]\\
  \hline
  B: Evaluate the faithfulness of explanations	& UAIA102: Users are not satisfied with the faithfulness of the explanation. [H9]	&UAIA105: The AI system does not evaluate the faithfulness of explanations. [H8]\\
  \hline
  C: Evaluate the monotonicity of explanations	& UAIA103: Users are not satisfied with the monotonicity of the explanation. [H11]	& UAIA106: The AI system does not evaluate the monotonicity of explanations. [H10]\\
  \botrule
  \end{tabular*} 
  \label{tab:list-ethical-UAIA}
\end{table*}

After UAIAs are identified, the causal scenarios for UAIAs need to be developed based on UAIAs according to the proposed framework (see Figure~\ref{fig:ethical-STPA}). Table~\ref{tab:list-causal-scenarios} provides examples of causal scenarios for UAIA101 (The AI system provides the wrong explanation) identified in the previous step.

\begin{table*}[!htb]
  \caption{List of causal scenarios for UAIA101 (The AI system provides the wrong explanation).}
  \begin{tabular*}{31pc}{@{}p{40pt}p{385pt}@{}}
  \hline
  ID & Causal scenarios \\
  \hline
  CS1	& The AI system used the wrong explanation method.\\
  CS2	& The explanation method used the wrong parameters.\\
  CS3	& The data used to generate explanations was not appropriate.\\
  CS4	& The system lacked documentation on the explanation method.\\
  \botrule
  \end{tabular*} 
  \label{tab:list-causal-scenarios}
\end{table*}

As shown in Figure~\ref{fig:ethical-STPA}, the developed causal scenarios for unethical AI actions are used to elicit ethical constraints and design recommendations. Table~\ref{tab:list-ethical-constraints} lists ethical constraints related to the hazards for transparency. Table~\ref{tab:list-ethical-constraints-design} shows examples of design recommendations for UAIA101 for the ethical constraint of EC101 (The AI system must not provide the wrong explanation).

  \begin{table*}[!htb]
    \caption{List of ethical constraints related to the ethical hazards for transparency.}
    \begin{tabular*}{31pc}{@{}p{60pt}p{25pt}p{260pt}@{}}
    \toprule
    Principle segment &Hazard ID & Ethical constraints \\
    \hline
    Traceability & H1 &  EC1: The information on the data, data collection, and data preprocessing should be documented.\\
    [3pt]         & & EC2: If the system loses this data documentation, it should be re-documented.\\
    [3pt] & H2 & EC3: The information on the method of training and testing the algorithm should be documented.\\
    [3pt] & & EC4: If the system loses this algorithm documentation, it should be re-documented.\\
    [3pt] & H3 & EC5: Documentation of the data used to test and validate the model is required.\\ 
      &  & EC6: If the system loses this test data documentation, it should be re-documented.\\
    [3pt] & H4 & EC7: The outcomes or decisions made by the algorithm must be documented.\\
    & & EC8: If the system loses this outcome/decision documentation, it should be re-documented.\\
    \hline
    Communication & H5 & EC9: The characteristics, limitations, and risks of the AI system should be communicated to users.\\
    &   & EC10: If the communication of characteristics, limitations, and risks of the AI system to end users is lost, it should be re-established.\\
    \hline
    Explainability & H6 & EC11: Explanations on the model or outputs should be provided.\\
    [3pt] & H7 & EC12: If explanations on the model or outputs are wrong, the explanations should be re-produced.\\
    [3pt] & H8 & EC13: The faithfulness of explanation should be provided.\\
    [3pt] & H9 & EC14: If users are not satisfied with the faithfulness of explanation, the explanation should be re-produced.\\
    [3pt] & H10 & EC15: The monotonicity of explanation should be provided.\\
    [3pt] & H11 & EC16: If users are not satisfied with the monotonicity of explanation, the explanation should be re-produced. \\
    \botrule
    \end{tabular*} 
    \label{tab:list-ethical-constraints}
  \end{table*}

  \begin{table*}[!htb]
    \caption{List of ethical constraints and design recommendation for UAIA101 (The AI system provides the wrong explanation).}
    \begin{tabular*}{31pc}{@{}p{30pt}p{325pt}@{}}
    \toprule
    ID & Ethical constraint/Design recommendation \\
    \hline
    EC101	& The AI system must not provide the wrong explanation.\\
    DR101.1	& System should require the confirmation that there are explanations provided.\\
    DR101.2	& System should provide the faithfulness of explanations.\\
    DR101.3	& The evaluation of faithfulness of explanations should use data with distributions similar to the distributions of actual input data.\\
    DR101.4	& System should provide documents on the method of explanation and method that is used to evaluate faithfulness of explanations.\\
    \botrule
    \end{tabular*} 
    \label{tab:list-ethical-constraints-design}
  \end{table*}

As profiled in Figure~\ref{fig:ethical-STPA}, the design recommendations are then further refined into ethical requirements for AI solutions. Table~\ref{tab:list-requirements} lists examples of ethical requirements elicited from design recommendations for UAIA101. These ethical requirements are used for the AI ethics assurance case. Table~\ref{tab:list-requirements} also provides examples of verification methods for ethical requirements, such as demonstration, black-box testing, scenario testing, and algorithmic evaluation.

  \begin{table*}[!htb]
    \caption{List of ethical requirements elicited from design recommendations for UAIA101 (The AI system provides the wrong explanation).}
    \begin{tabular*}{31pc}{@{}p{30pt}p{170pt}p{60pt}p{67pt}@{}}
    \toprule
    ID	& Ethical requirement	& Traceability to DRs	& Verification method\\
    \hline
    R101.1	& System should display explanations besides the AI outcomes/decisions.	& DR101.1	& Demonstration\\
    R101.2	& System should display the faithfulness of explanations.	& DR101.2	& Algorithmic evaluation\\
    R101.3	& System should confirm that the distribution of data used for the evaluation of faithfulness of explanations is similar to distributions of actual input data.	& DR101.3	& Black-box testing\\
    R101.4	& System should document the method for the explanation and evaluation of faithfulness of explanations.	& DR101.4	& Demonstration\\
    \botrule
    \end{tabular*} 
    \label{tab:list-requirements}
    \end{table*}

\begin{figure*}[!htb]
    \centering
    \includegraphics[width=0.99\linewidth]{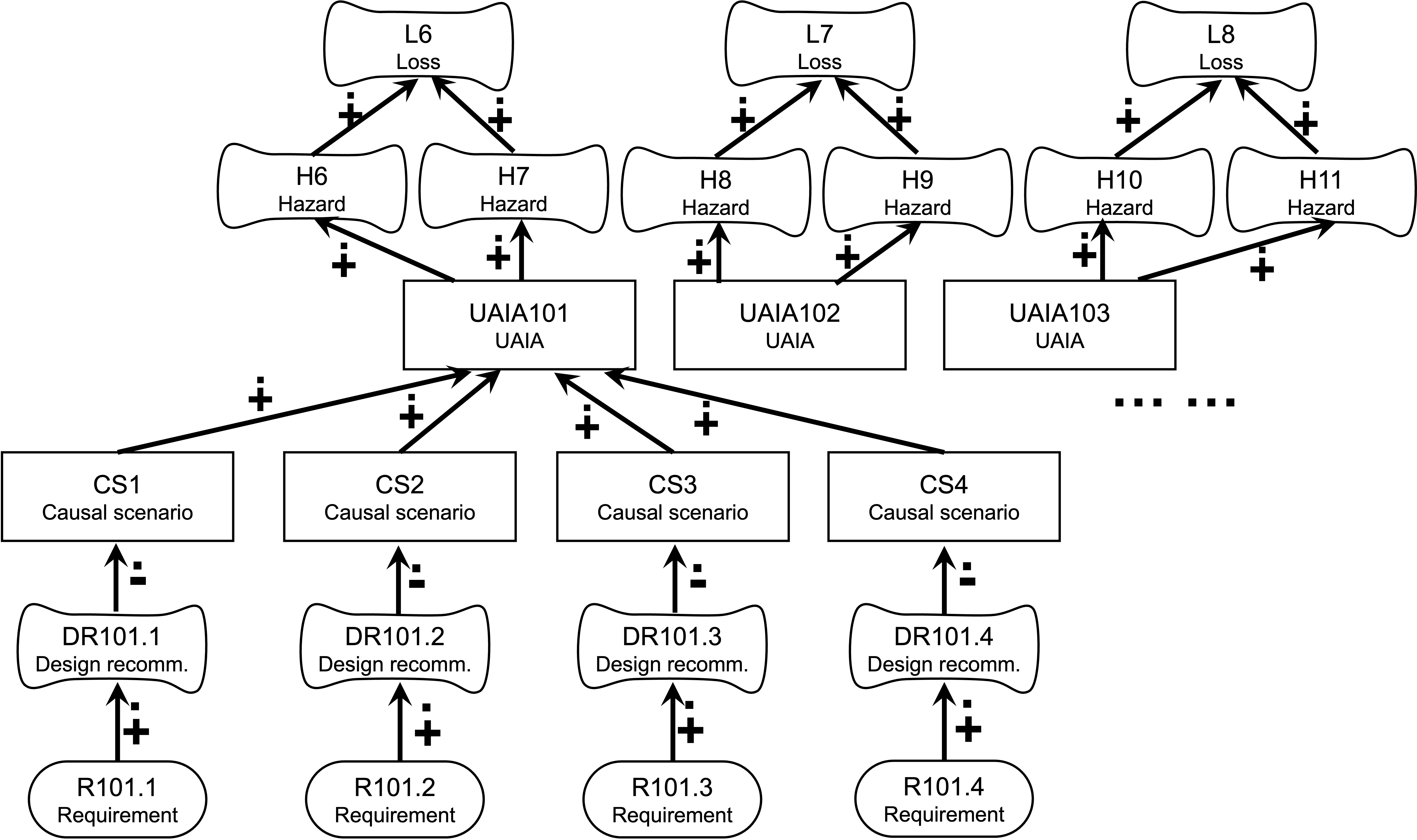}
      \caption{The AI ethics assurance case for the explainability.}
  \label{fig:assurance-case-explainability}       
  \end{figure*}

Based on these previous steps from different ethical states identification to the identification of ethical requirements for an AI solution, a URN-based assurance case for the AI ethics is created. Figure~\ref{fig:assurance-case-explainability} shows an example of assurance case for the explainability. This assurance case focuses on the ethical risks that the AI solution may expose through the ethical requirements. From GRL representations of Figure~\ref{fig:assurance-case-explainability}, users can visually assess the ethical issues following the graph and confirm that ``all ethical hazards are mitigated''. Rather than arguing that the system is free from ethical hazards, as many previous efforts have done, the focus is on addressing user requirements related to ethical risks and providing evidence to mitigate those risks.

\section{E-LENS: An extensible platform for AI ethics assurance cases}

In order to perform the AI ethics assurance case as shown in Figure~\ref{fig:assurance-case-explainability} easily by users, this paper proposes a platform named E-LENS (Ethical-Lens) and incorporates AI ethical requirements into this platform to implement the user requirements-oriented AI ethics assurance. This platform is designed to be extensible so that new components, such as new ethical principles and ethical questions related to ethical requirements, can be added if necessary.

E-LENS has the following modules including verification methods and other features (see Figure~\ref{fig:elens_components}):

\begin{figure}[!htb]
  \centering
  \includegraphics[width=0.5\linewidth]{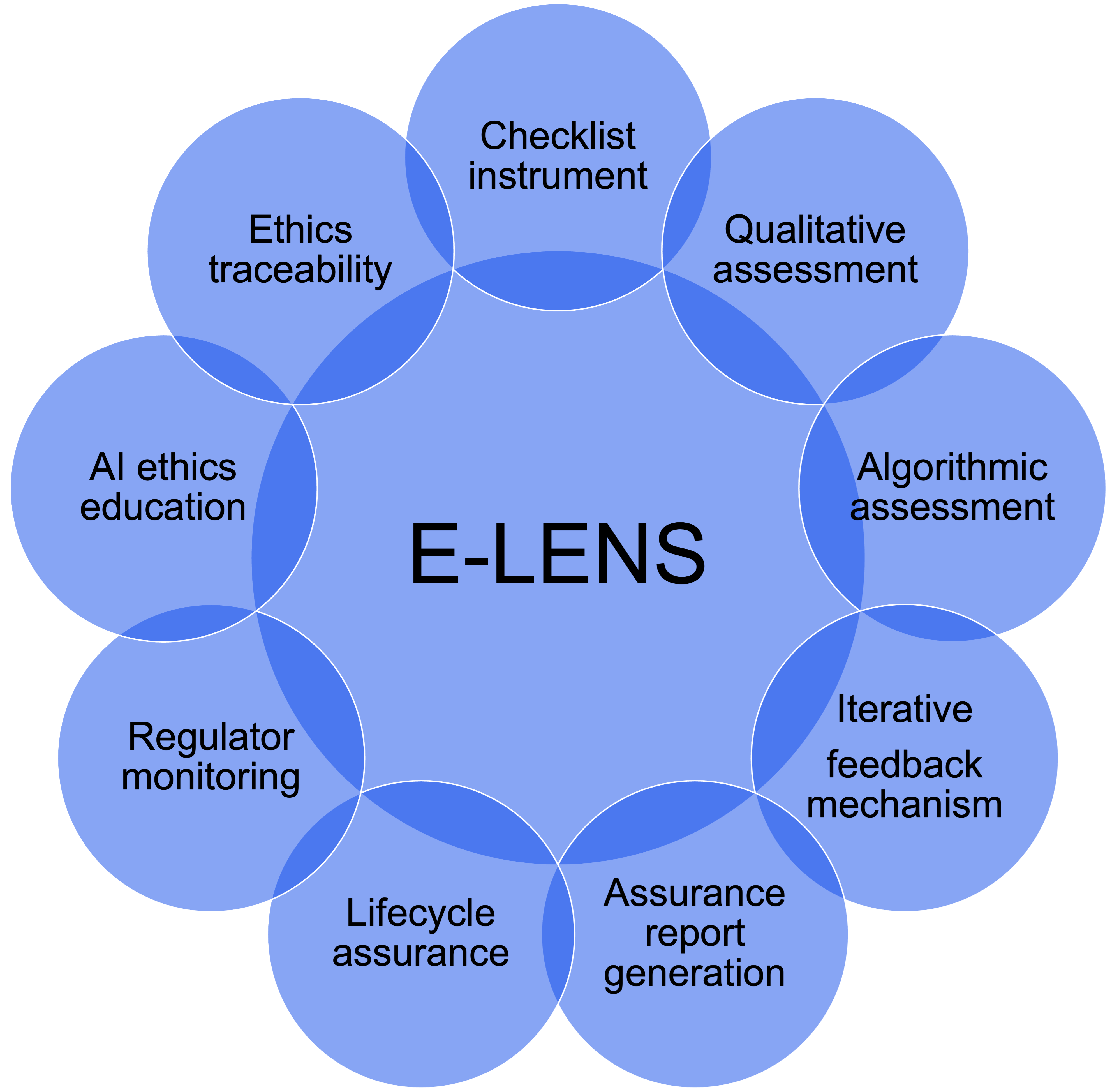}
    \caption{Modules included in the E-LENS framework.}
\label{fig:elens_components}       
\end{figure}

\begin{itemize}
    \item \textbf{Checklist instrument}: A checklist of questionnaire, which corresponds to ethical requirements derived in the assurance case, helps users to easily check each ethical requirement and provide evidences for the assurance.
    \item \textbf{Qualitative assessment}: Considering the nature of ethics, qualitative approaches are used to assess ethical requirements.
    \item \textbf{Algorithmic assessment}: Given the data-driven nature of machine learning and AI, algorithmic approaches are used to answer ethical requirements in the assurance.
    \item \textbf{Iterative feedback mechanism}: The AI ethics assurance is designed as an iterative process. AI solution providers and ethics validators communicate in an iterative feedback loop to make sure all ethics requirements are met.
    \item \textbf{Assurance report generation}: The assurance report documents the overall AI ethical requirements and their evidences or answers from AI solution providers, which allows the easy communication among stakeholders regarding AI ethics.
    \item \textbf{Lifecycle assurance}: The assurance encompasses the entire AI lifecycle, ensuring ethical safety at every stage.
    \item \textbf{Regulator monitoring}: E-LENS introduces the regulator into the AI ethics assurance pipeline and allows the AI ethics monitoring in real-time by regulators.
    \item \textbf{AI ethics education}: Education on AI ethics helps stakeholders better understand ethical principles, leading to more effective implementation of those principles in AI systems.
    \item \textbf{Ethics traceability}: The evidence for each AI ethics argument at every AI lifecycle stage is recorded in E-LENS, which allows to access and trace AI ethics status at any time.
\end{itemize}

\begin{figure*}[!htb]
  \centering
  \includegraphics[width=0.99\linewidth]{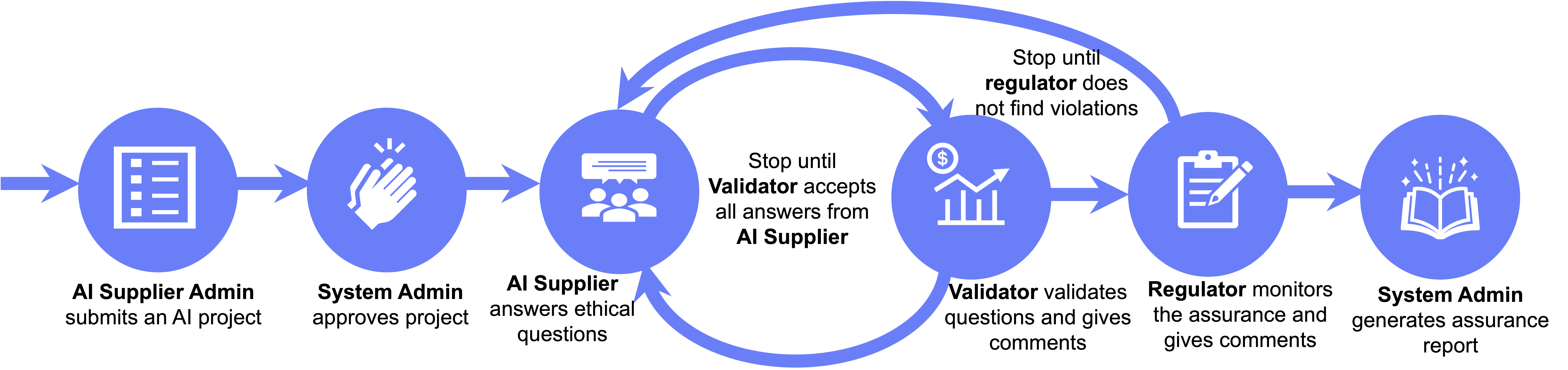}
    \caption{The pipeline of the AI ethics assurance in E-LENS.}
\label{fig:pipeline}       
\end{figure*}

Each of these features are introduced in more details in the following subsections. Figure~\ref{fig:pipeline} shows the pipeline of the AI ethics assurance in E-LENS.

\subsection{Checklist instrument}

Despite the theoretical weaknesses in its framework, the checklist-style questionnaires are widely employed as the standard ethics governance model in biomedical and clinical research \cite{canca2019new}. Similar to checklist-style questionnaires employed in the biomedical and clinical research, our work proposes using a checklist as a tool to validate the ethics surrounding AI solutions. It is ideal to develop checklist-style questionnaires for each identified ethical principle to audit AI compliance with these principles.



\begin{figure*}[!htb]
  \centering
  \includegraphics[width=0.9\linewidth]{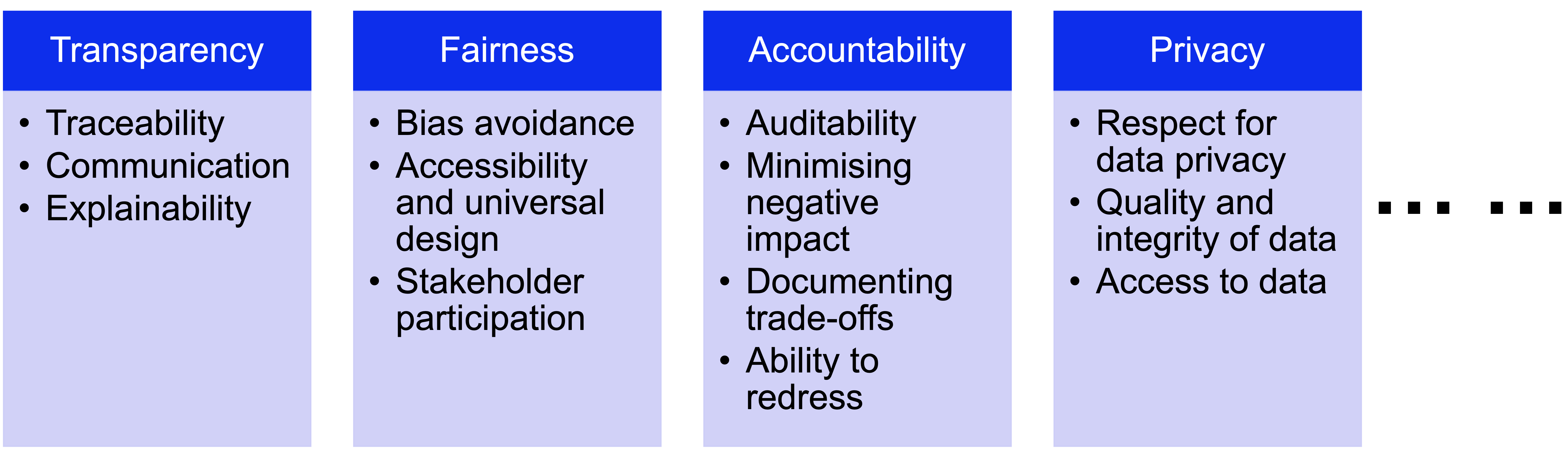}
    \caption{Ethical principles and their segments.}
\label{fig:ethical_principles}       
\end{figure*}

Inspired by the assessment list for trustworthy AI published by the HLEG-AI of the European Commission \cite{ala2020assessment}, this paper develops a hierarchical checklist questionnaire approach for ethical requirements to assure AI ethical principles. Each ethical principle has an overarching question on ethical requirements which is split into different sub-questions from different perspectives (segments). For example, the transparency principle is evaluated from the perspectives (segments) of requirements on traceability, communication, and explainability (see Figure~\ref{fig:ethical_principles}).     
The previous survey indicates that ethical principles for AI commonly converge on transparency, justice and fairness, responsibility and accountability, non-maleficence, privacy, beneficence, freedom and autonomy, trust, sustainability, dignity, and solidarity \cite{zhou2020survey}. This paper focuses on mandatory ethical principles of transparency, fairness, accountability and privacy. 
We categorise the existing ethical principle questions into different requirements for different AI lifecycle stages and desiderata \cite{ashmore2021assuring}: relevant, complete, balanced, and accurate. Furthermore, the hierarchical checklist approach is designed to be extensible in E-LENS, which allows users to add new ethical principles, new principle segments and new checklist questions if necessary in the E-LENS framework.

\subsection{Qualitative assessment}

Ethics and AI are two distinct scientific domains. Ethics is a highly abstract social science, whereas AI involves defining strict quantitative metrics, such as prediction accuracy. Therefore, we propose implementing AI ethical principles both qualitatively and quantitatively/algorithmically. 

The qualitative assessment is designed as checklist questions that are answered with qualitative descriptions. In the E-LENS framework, qualitative questions are classified into two types: multiple choice questions and extended response questions. Multiple choice questions ask users to select one answer from provided options, while extended response questions require users to write short descriptions as answers on how AI solutions solve the ethical issues. For example, extended response questions are used to validate the measures implemented for protecting data privacy.

\subsection{Algorithmic assessment}

Quantitative or algorithmic assessment aims to evaluate how well AI solutions adhere to ethical principles by developing algorithmic approaches for this purpose. For instance, quantitative measures can be used to validate the explainability of AI solutions and assess the fairness of both data and models. 
Various approaches can be used to enhance the transparency of an AI model or algorithm \cite{zhou2021evaluating}, assess algorithmic bias for fairness in predictions \cite{Bellamy_fairness_2018}, or improve privacy through algorithms \cite{kairouz2021advances}. These algorithmic methods can be effective for AI experts and developers in detecting and assessing ethical issues within AI models.

In the E-LENS framework, algorithms are developed to evaluate the AI model explainability as well as fairness of data and AI models. They require users to provide relevant data and model information for the evaluation.

\subsection{Feedback mechanism}

The AI ethics assurance process is designed as an iterative looping process, where AI suppliers answer ethical questions and the answers are then evaluated by AI ethics validators (see Figure~\ref{fig:pipeline}). AI ethics validators give feedback to AI suppliers for any further improvements until AI ethics validators accept all answers from AI suppliers.

\subsection{Assurance report generation}

AI ethics assurance report documents the ethics status and support information on AI solutions. It helps to communicate among stakeholders on AI ethics. In the E-LENS framework, all answers to ethics questions are saved in a database. A report generation function is developed to create PDF documents to stakeholders on the AI ethics status and support information. Two types of report are provided in the E-LENS framework: 1) the full report which documents all details of answers including comments from AI ethics validators to all AI ethics questions; and 2) the summary report which provides short summary for the ethics status of each ethical principles.

\subsection{Lifecycle assurance}

\begin{figure*}[!htb]
  \centering
  \includegraphics[width=0.99\linewidth]{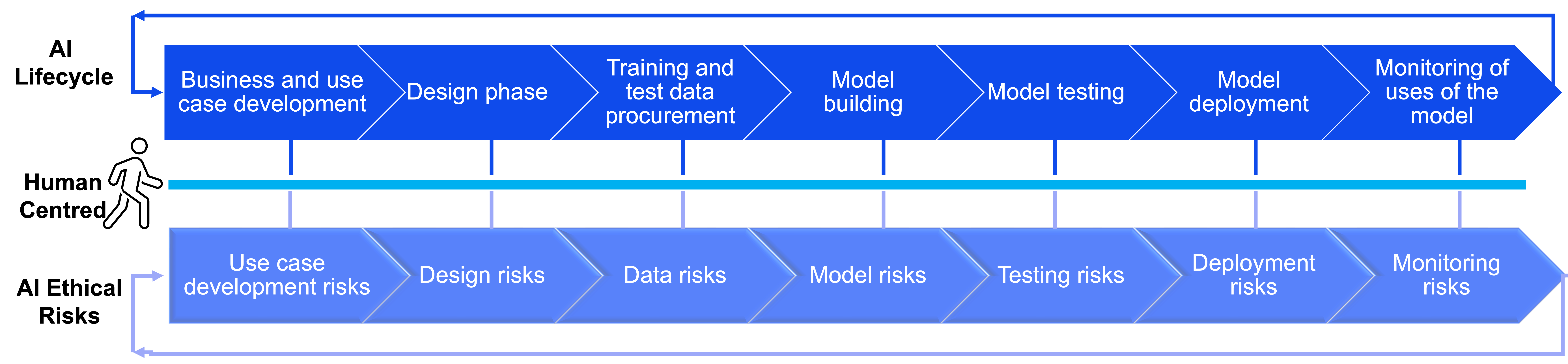}
    \caption{The AI lifecycle and ethical risks.}
\label{fig:lifecycle_risk}       
\end{figure*}

The lifecycle of a typical AI application typically encompasses several stages: business and use-case development, design, data collection for training and testing, model building and testing, model deployment, and system monitoring (see Figure~\ref{fig:lifecycle_risk}). This framework provides a high-level view of how an AI project should be structured to deliver practical business value at each stage. Morley et al. \cite{morley2020initial} emphasize that ethical principles should be integrated at every stage of the AI lifecycle to ensure that the AI system is designed, implemented, and deployed ethically.

In the E-LENS framework, checklist questions are designed for every ethical principle at each stage to make sure that the AI ethics is assured covering the AI lifecycle.

\subsection{Regulator monitoring}

Although it is not fully established in many countries, the regulation framework for AI ethics plays an important role in the success of assurance of AI ethics. In the E-LENS framework, regulators are introduced into the assurance pipeline of AI ethics as shown in Figure~\ref{fig:pipeline}. In this pipeline, after the validator finishes the assessment, the regulator checks the assessment results to monitor whether there are any violations to related regulations. If yes, the assurance process returns to the previous step and the AI supplier is required to fix any comments from the regulator for further assessment until the regulator does not find further violations to related regulations (see Figure~\ref{fig:pipeline}).

\subsection{Ethics traceability}

Ethics traceability refers to the capability that relevant stakeholders can trace back to the previous ethics status or support evidence for ethical principles. It allows stakeholders easily learn ethical issues and make informed decisions. In the E-LENS framework, all of the ethics status and support evidence are saved in databases, stakeholders can easily access the information to trace back the assurance of AI ethics for AI solutions. 


\section{AI ETHICS ASSURANCE BEST PRACTICE: AI FOR RECRUITMENT}

This paper presents the E-LENS framework for assuring ethical principles in AI, implemented as a web-based application to effectively ensure ethical compliance in AI solutions. Designed to accommodate the evolving nature of ethical investigations, the platform is open and flexible, allowing for the addition or customization of ethical principles and corresponding qualitative and quantitative questions to suit various project requirements. Upon completion of the validation process, the platform generates a summary report for AI solution providers, detailing the ethical aspects addressed and identifying areas for improvement from an ethical standpoint. In this section, we show a case study of the use of E-LENS in the AI ethics assurance.

In the human resources (HR) recruiting process, one of the most challenge steps is to shortlist a number of most qualified applications for a job position for interviews by reviewing each application \cite{gusdorf2008recruitment}. A large number of applications (even hundreds or thousands) are submitted if the posted position is popular. It is much time consuming if all applications are reviewed by humans one by one. Even more, different reviewers may have different comments on the same application, which cause further bias in the selection. Each application may include resume, cover letter, answers to criteria, certificates and other supporting documents. HR departments often have datasets that include potential job candidates, past applicants, and current employees. AI is used to generate analytics-backed insights into various HR processes, including the shortlisting of candidates during recruitment \cite{kaushal2023artificial,roopalatha2024artificial}.

However, AI may raise myriad ethical issues because of the use of human related data. Some of typical ethical issues include transparency, data privacy, gender bias, and/or unclear accountability \cite{ajonbadi2024before}. The assessment of the AI algorithm's adherence to core ethical principles is therefore critical to delivering a responsible solution which proactively addresses potential risk. 

\begin{figure*}[!htb]
  \centering
  \includegraphics[width=0.9\linewidth]{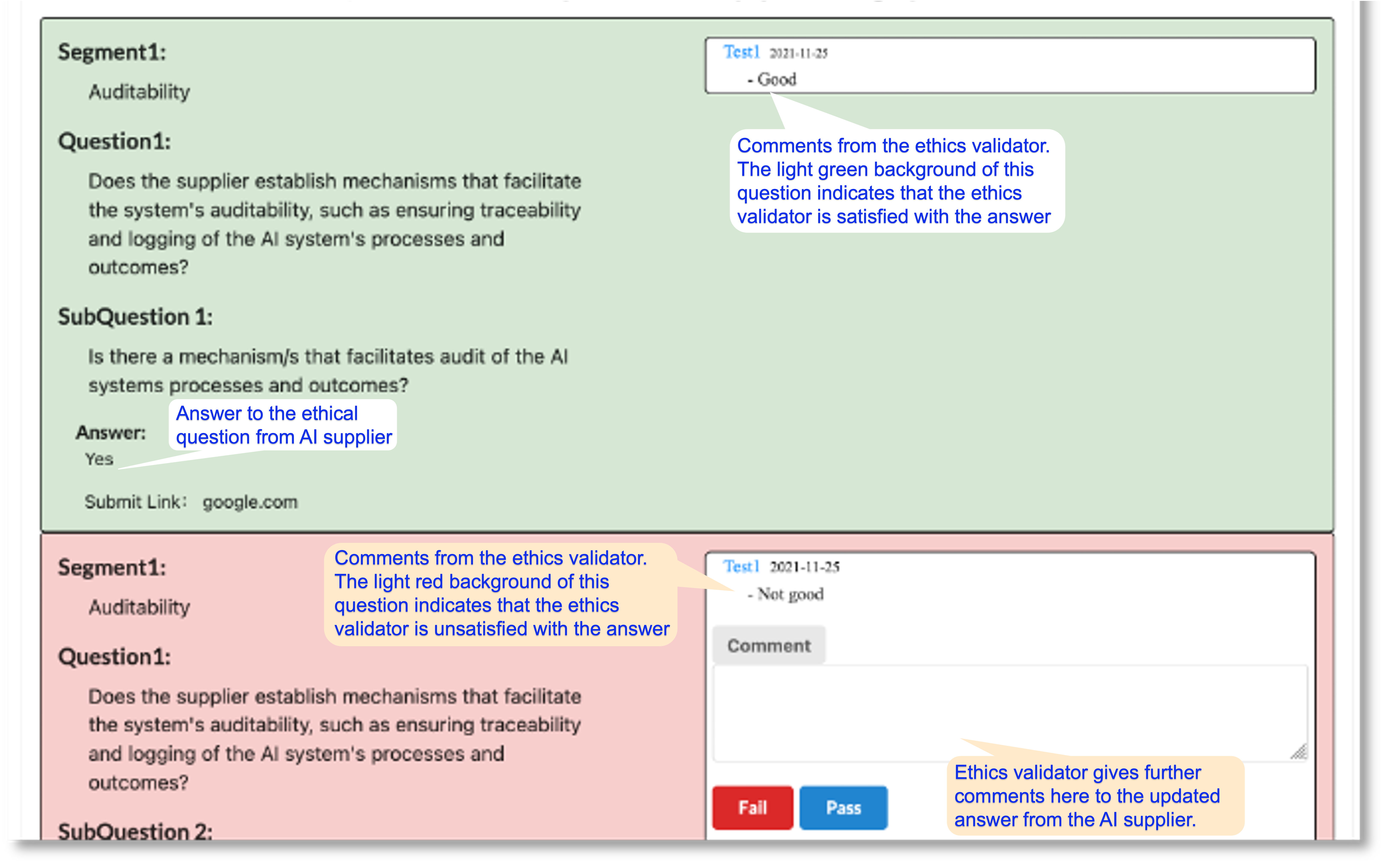}
    \caption{The feedback mechanism implemented in E-LENS.}
\label{fig:feedback-in-elens}       
\end{figure*}

In this section, the proposed user requirements-based E-LENS platform is applied to assure AI ethics of an AI-driven HR shortlisting system. Firstly, user requirements are defined from the ethical perspective for the shortlisting system such as free of bias to a certain gender and transparent in the shortlisting decision making. These ethical requirements act as the basis for the setup of the ethics assurance case. Following this step, a series of checklist-style questionnaires are used to gather evidence for assessing the ethics of AI solutions both qualitatively and quantitatively. For instance, qualitative questions validate measures used to protect data privacy, while quantitative questions evaluate AI compliance with ethical principles through specific metrics. Examples of quantitative measures include assessing the faithfulness (the relevance of features) and monotonicity (the effect of individual features on model performance) of AI explainability, as well as examining the fairness of both data and AI models.
Ethics validators and regulators then check the answers to ethical questions to assure answers. As shown in Figure~\ref{fig:pipeline}, this is a looping process until validators and regulators are satisfied with the evidence as answers to questions. Figure~\ref{fig:feedback-in-elens} shows such feedback looping mechanism implemented in E-LENS. In the E-LENS implementation, the ethics validator/regulator gives comments to the answer to the ethics question from the AI supplier. If he/she is satisfied with the answer, the background of this question will become light green and no further comments can be given after he/she submits comments. Otherwise, the background of this question will be light red and the ethics validator/regulator is allowed to given further comments to further answers until he/she is satisfied with the answer. Finally, the AI ethics assurance report is generated as a PDF file based on answers to ethical questions which are relevant to user ethical requirements.






\section{CONCLUSION AND FUTURE WORK}

This paper introduces the concept of assurance cases, which is widely used in safety-critical systems for the safety assurance, into the AI ethics assurance for a new approach of user requirements-oriented AI ethics assurance. The new approach addresses the significant challenge of assurance of AI ethics by integrating user requirements, evidence, and validation into the AI ethics assurance case. This paper provides a robust solution that uses the widely accepted safety assurance theory in solving AI ethics assurances. The implementation of E-LENS for the user requirements-oriented AI ethics assurance cases allows the versatile and flexible assurance of AI ethics in practices. 
The future work of this paper will focus on the development of AI ethics assurance cases based on the proposed user requirements-oriented approach for specific domains or applications and integrate them into the E-LENS platform. Ideally, assurance case development occurs in parallel to the AI solution development, resulting in a legitimately strong argument for the desired ethical requirement.

\backmatter

\section*{Acknowledgements}

\section*{Author contributions}
All authors have contributed to the formation, discussion, and writing of this article.

\section*{Competing interests} 
The authors declare no competing interests.

\section*{Data availability} 
Data are contained within the article.

\bibliography{references,ai_ethics_assurance}


\begin{thebibliography}{53}
\ifx \bisbn   \undefined \def \bisbn  #1{ISBN #1}\fi
\ifx \binits  \undefined \def \binits#1{#1}\fi
\ifx \bauthor  \undefined \def \bauthor#1{#1}\fi
\ifx \batitle  \undefined \def \batitle#1{#1}\fi
\ifx \bjtitle  \undefined \def \bjtitle#1{#1}\fi
\ifx \bvolume  \undefined \def \bvolume#1{\textbf{#1}}\fi
\ifx \byear  \undefined \def \byear#1{#1}\fi
\ifx \bissue  \undefined \def \bissue#1{#1}\fi
\ifx \bfpage  \undefined \def \bfpage#1{#1}\fi
\ifx \blpage  \undefined \def \blpage #1{#1}\fi
\ifx \burl  \undefined \def \burl#1{\textsf{#1}}\fi
\ifx \doiurl  \undefined \def \doiurl#1{\url{https://doi.org/#1}}\fi
\ifx \betal  \undefined \def \betal{\textit{et al.}}\fi
\ifx \binstitute  \undefined \def \binstitute#1{#1}\fi
\ifx \binstitutionaled  \undefined \def \binstitutionaled#1{#1}\fi
\ifx \bctitle  \undefined \def \bctitle#1{#1}\fi
\ifx \beditor  \undefined \def \beditor#1{#1}\fi
\ifx \bpublisher  \undefined \def \bpublisher#1{#1}\fi
\ifx \bbtitle  \undefined \def \bbtitle#1{#1}\fi
\ifx \bedition  \undefined \def \bedition#1{#1}\fi
\ifx \bseriesno  \undefined \def \bseriesno#1{#1}\fi
\ifx \blocation  \undefined \def \blocation#1{#1}\fi
\ifx \bsertitle  \undefined \def \bsertitle#1{#1}\fi
\ifx \bsnm \undefined \def \bsnm#1{#1}\fi
\ifx \bsuffix \undefined \def \bsuffix#1{#1}\fi
\ifx \bparticle \undefined \def \bparticle#1{#1}\fi
\ifx \barticle \undefined \def \barticle#1{#1}\fi
\bibcommenthead
\ifx \bconfdate \undefined \def \bconfdate #1{#1}\fi
\ifx \botherref \undefined \def \botherref #1{#1}\fi
\ifx \url \undefined \def \url#1{\textsf{#1}}\fi
\ifx \bchapter \undefined \def \bchapter#1{#1}\fi
\ifx \bbook \undefined \def \bbook#1{#1}\fi
\ifx \bcomment \undefined \def \bcomment#1{#1}\fi
\ifx \oauthor \undefined \def \oauthor#1{#1}\fi
\ifx \citeauthoryear \undefined \def \citeauthoryear#1{#1}\fi
\ifx \endbibitem  \undefined \def \endbibitem {}\fi
\ifx \bconflocation  \undefined \def \bconflocation#1{#1}\fi
\ifx \arxivurl  \undefined \def \arxivurl#1{\textsf{#1}}\fi
\csname PreBibitemsHook\endcsname

\bibitem[\protect\citeauthoryear{Vinuesa et~al.}{2020}]{vinuesa2020role}
\begin{barticle}
\bauthor{\bsnm{Vinuesa}, \binits{R.}},
\bauthor{\bsnm{Azizpour}, \binits{H.}},
\bauthor{\bsnm{Leite}, \binits{I.}},
\bauthor{\bsnm{Balaam}, \binits{M.}},
\bauthor{\bsnm{Dignum}, \binits{V.}},
\bauthor{\bsnm{Domisch}, \binits{S.}},
\bauthor{\bsnm{Fell{\"a}nder}, \binits{A.}},
\bauthor{\bsnm{Langhans}, \binits{S.D.}},
\bauthor{\bsnm{Tegmark}, \binits{M.}},
\bauthor{\bsnm{Fuso~Nerini}, \binits{F.}}:
\batitle{The role of artificial intelligence in achieving the sustainable development goals}.
\bjtitle{Nature communications}
\bvolume{11}(\bissue{1}),
\bfpage{1}--\blpage{10}
(\byear{2020})
\end{barticle}
\endbibitem

\bibitem[\protect\citeauthoryear{Zhou and Chen}{2023}]{zhou2023ai}
\begin{barticle}
\bauthor{\bsnm{Zhou}, \binits{J.}},
\bauthor{\bsnm{Chen}, \binits{F.}}:
\batitle{Ai ethics: From principles to practice}.
\bjtitle{AI and SOCIETY}
\bvolume{38}(\bissue{6}),
\bfpage{2693}--\blpage{2703}
(\byear{2023})
\end{barticle}
\endbibitem

\bibitem[\protect\citeauthoryear{Zhou et~al.}{2020}]{zhou2020survey}
\begin{bchapter}
\bauthor{\bsnm{Zhou}, \binits{J.}},
\bauthor{\bsnm{Chen}, \binits{F.}},
\bauthor{\bsnm{Berry}, \binits{A.}},
\bauthor{\bsnm{Reed}, \binits{M.}},
\bauthor{\bsnm{Zhang}, \binits{S.}},
\bauthor{\bsnm{Savage}, \binits{S.}}:
\bctitle{A survey on ethical principles of ai and implementations}.
In: \bbtitle{2020 IEEE Symposium Series on Computational Intelligence (SSCI)},
pp. \bfpage{3010}--\blpage{3017}
(\byear{2020}).
\bcomment{IEEE}
\end{bchapter}
\endbibitem

\bibitem[\protect\citeauthoryear{Dawson et~al.}{2019}]{dawson2019artificial}
\begin{botherref}
\oauthor{\bsnm{Dawson}, \binits{D.}},
\oauthor{\bsnm{Schleiger}, \binits{E.}},
\oauthor{\bsnm{Horton}, \binits{J.}},
\oauthor{\bsnm{McLaughlin}, \binits{J.}},
\oauthor{\bsnm{Robinson}, \binits{C.}},
\oauthor{\bsnm{Quezada}, \binits{G.}},
\oauthor{\bsnm{Scowcroft}, \binits{J.}},
\oauthor{\bsnm{Hajkowicz}, \binits{S.}}:
Artificial intelligence: Australia’s ethics framework-a discussion paper
(2019)
\end{botherref}
\endbibitem

\bibitem[\protect\citeauthoryear{Chatila and Havens}{2019}]{chatila2019ieee}
\begin{botherref}
\oauthor{\bsnm{Chatila}, \binits{R.}},
\oauthor{\bsnm{Havens}, \binits{J.C.}}:
The ieee global initiative on ethics of autonomous and intelligent systems.
Robotics and well-being,
11--16
(2019)
\end{botherref}
\endbibitem

\bibitem[\protect\citeauthoryear{Jobin et~al.}{2019}]{jobin2019global}
\begin{barticle}
\bauthor{\bsnm{Jobin}, \binits{A.}},
\bauthor{\bsnm{Ienca}, \binits{M.}},
\bauthor{\bsnm{Vayena}, \binits{E.}}:
\batitle{The global landscape of ai ethics guidelines}.
\bjtitle{Nature Machine Intelligence}
\bvolume{1}(\bissue{9}),
\bfpage{389}--\blpage{399}
(\byear{2019})
\end{barticle}
\endbibitem

\bibitem[\protect\citeauthoryear{{Digital.NSW}}{2024}]{digitalnsw2024}
\begin{botherref}
\oauthor{\bsnm{{Digital.NSW}}}:
Mandatory Ethical Principles for the use of {AI}.
\url{https://www.digital.nsw.gov.au/policy/artificial-intelligence/artificial-intelligence-ethics-policy/mandatory-ethical-principles}
(2024)
\end{botherref}
\endbibitem

\bibitem[\protect\citeauthoryear{B{\'e}lisle-Pipon et~al.}{2022}]{belisle2022artificial}
\begin{botherref}
\oauthor{\bsnm{B{\'e}lisle-Pipon}, \binits{J.-C.}},
\oauthor{\bsnm{Monteferrante}, \binits{E.}},
\oauthor{\bsnm{Roy}, \binits{M.-C.}},
\oauthor{\bsnm{Couture}, \binits{V.}}:
Artificial intelligence ethics has a black box problem.
AI \& SOCIETY,
1--16
(2022)
\end{botherref}
\endbibitem

\bibitem[\protect\citeauthoryear{Zicari et~al.}{2022}]{zicari2022assess}
\begin{botherref}
\oauthor{\bsnm{Zicari}, \binits{R.V.}},
\oauthor{\bsnm{Amann}, \binits{J.}},
\oauthor{\bsnm{Bruneault}, \binits{F.}},
\oauthor{\bsnm{Coffee}, \binits{M.}},
\oauthor{\bsnm{D{\"u}dder}, \binits{B.}},
\oauthor{\bsnm{Hickman}, \binits{E.}},
\oauthor{\bsnm{Gallucci}, \binits{A.}},
\oauthor{\bsnm{Gilbert}, \binits{T.K.}},
\oauthor{\bsnm{Hagendorff}, \binits{T.}},
\oauthor{\bsnm{Halem}, \binits{I.}}, et al.:
How to assess trustworthy ai in practice.
arXiv preprint arXiv:2206.09887
(2022)
\end{botherref}
\endbibitem

\bibitem[\protect\citeauthoryear{Ala-Pietil{\"a} et~al.}{2020}]{ala2020assessment}
\begin{bbook}
\bauthor{\bsnm{Ala-Pietil{\"a}}, \binits{P.}},
\bauthor{\bsnm{Bonnet}, \binits{Y.}},
\bauthor{\bsnm{Bergmann}, \binits{U.}},
\bauthor{\bsnm{Bielikova}, \binits{M.}},
\bauthor{\bsnm{Bonefeld-Dahl}, \binits{C.}},
\bauthor{\bsnm{Bauer}, \binits{W.}},
\bauthor{\bsnm{Bouarfa}, \binits{L.}},
\bauthor{\bsnm{Chatila}, \binits{R.}},
\bauthor{\bsnm{Coeckelbergh}, \binits{M.}},
\bauthor{\bsnm{Dignum}, \binits{V.}}, \betal:
\bbtitle{The Assessment List for Trustworthy Artificial Intelligence (ALTAI)}.
\bpublisher{European Commission},
\blocation{Brussels, Belgium}
(\byear{2020})
\end{bbook}
\endbibitem

\bibitem[\protect\citeauthoryear{Han and Choi}{2022}]{han2022checklist}
\begin{bchapter}
\bauthor{\bsnm{Han}, \binits{S.-H.}},
\bauthor{\bsnm{Choi}, \binits{H.-J.}}:
\bctitle{Checklist for validating trustworthy ai}.
In: \bbtitle{2022 IEEE International Conference on Big Data and Smart Computing (BigComp)},
pp. \bfpage{391}--\blpage{394}
(\byear{2022}).
\bcomment{IEEE}
\end{bchapter}
\endbibitem

\bibitem[\protect\citeauthoryear{Gebru et~al.}{2021}]{gebru2021datasheets}
\begin{barticle}
\bauthor{\bsnm{Gebru}, \binits{T.}},
\bauthor{\bsnm{Morgenstern}, \binits{J.}},
\bauthor{\bsnm{Vecchione}, \binits{B.}},
\bauthor{\bsnm{Vaughan}, \binits{J.W.}},
\bauthor{\bsnm{Wallach}, \binits{H.}},
\bauthor{\bsnm{Iii}, \binits{H.D.}},
\bauthor{\bsnm{Crawford}, \binits{K.}}:
\batitle{Datasheets for datasets}.
\bjtitle{Communications of the ACM}
\bvolume{64}(\bissue{12}),
\bfpage{86}--\blpage{92}
(\byear{2021})
\end{barticle}
\endbibitem

\bibitem[\protect\citeauthoryear{Arnold et~al.}{2019}]{arnold2019factsheets}
\begin{barticle}
\bauthor{\bsnm{Arnold}, \binits{M.}},
\bauthor{\bsnm{Bellamy}, \binits{R.K.}},
\bauthor{\bsnm{Hind}, \binits{M.}},
\bauthor{\bsnm{Houde}, \binits{S.}},
\bauthor{\bsnm{Mehta}, \binits{S.}},
\bauthor{\bsnm{Mojsilovi{\'c}}, \binits{A.}},
\bauthor{\bsnm{Nair}, \binits{R.}},
\bauthor{\bsnm{Ramamurthy}, \binits{K.N.}},
\bauthor{\bsnm{Olteanu}, \binits{A.}},
\bauthor{\bsnm{Piorkowski}, \binits{D.}}, \betal:
\batitle{Factsheets: Increasing trust in ai services through supplier's declarations of conformity}.
\bjtitle{IBM Journal of Research and Development}
\bvolume{63}(\bissue{4/5}),
\bfpage{6}--\blpage{1}
(\byear{2019})
\end{barticle}
\endbibitem

\bibitem[\protect\citeauthoryear{Adkins et~al.}{2022}]{adkins2022method}
\begin{bchapter}
\bauthor{\bsnm{Adkins}, \binits{D.}},
\bauthor{\bsnm{Alsallakh}, \binits{B.}},
\bauthor{\bsnm{Cheema}, \binits{A.}},
\bauthor{\bsnm{Kokhlikyan}, \binits{N.}},
\bauthor{\bsnm{McReynolds}, \binits{E.}},
\bauthor{\bsnm{Mishra}, \binits{P.}},
\bauthor{\bsnm{Procope}, \binits{C.}},
\bauthor{\bsnm{Sawruk}, \binits{J.}},
\bauthor{\bsnm{Wang}, \binits{E.}},
\bauthor{\bsnm{Zvyagina}, \binits{P.}}:
\bctitle{Method cards for prescriptive machine-learning transparency}.
In: \bbtitle{Proceedings of the 1st International Conference on AI Engineering: Software Engineering for AI},
pp. \bfpage{90}--\blpage{100}
(\byear{2022})
\end{bchapter}
\endbibitem

\bibitem[\protect\citeauthoryear{Zicari et~al.}{2021}]{zicari2021z}
\begin{barticle}
\bauthor{\bsnm{Zicari}, \binits{R.V.}},
\bauthor{\bsnm{Brodersen}, \binits{J.}},
\bauthor{\bsnm{Brusseau}, \binits{J.}},
\bauthor{\bsnm{D{\"u}dder}, \binits{B.}},
\bauthor{\bsnm{Eichhorn}, \binits{T.}},
\bauthor{\bsnm{Ivanov}, \binits{T.}},
\bauthor{\bsnm{Kararigas}, \binits{G.}},
\bauthor{\bsnm{Kringen}, \binits{P.}},
\bauthor{\bsnm{McCullough}, \binits{M.}},
\bauthor{\bsnm{M{\"o}slein}, \binits{F.}}, \betal:
\batitle{Z-inspection{\textregistered}: a process to assess trustworthy ai}.
\bjtitle{IEEE Transactions on Technology and Society}
\bvolume{2}(\bissue{2}),
\bfpage{83}--\blpage{97}
(\byear{2021})
\end{barticle}
\endbibitem

\bibitem[\protect\citeauthoryear{Rinehart et~al.}{2015}]{rinehart2015current}
\begin{botherref}
\oauthor{\bsnm{Rinehart}, \binits{D.J.}},
\oauthor{\bsnm{Knight}, \binits{J.C.}},
\oauthor{\bsnm{Rowanhill}, \binits{J.}}:
Current practices in constructing and evaluating assurance cases with applications to aviation.
Technical report,
NASA
(2015).
https://ntrs.nasa.gov/search.jsp?R=20150002819
\end{botherref}
\endbibitem

\bibitem[\protect\citeauthoryear{Sharifi et~al.}{2022}]{sharifi2022towards}
\begin{bchapter}
\bauthor{\bsnm{Sharifi}, \binits{S.}},
\bauthor{\bsnm{Amyot}, \binits{D.}},
\bauthor{\bsnm{Mylopoulos}, \binits{J.}},
\bauthor{\bsnm{McLaughlin}, \binits{P.}},
\bauthor{\bsnm{Feodoroff}, \binits{R.}}:
\bctitle{Towards improved certification of complex fintech systems--a requirements-based approach}.
In: \bbtitle{2022 IEEE 30th International Requirements Engineering Conference Workshops (REW)},
pp. \bfpage{205}--\blpage{214}
(\byear{2022}).
\bcomment{IEEE}
\end{bchapter}
\endbibitem

\bibitem[\protect\citeauthoryear{Picardi et~al.}{2020}]{picardi2020assurance}
\begin{bchapter}
\bauthor{\bsnm{Picardi}, \binits{C.}},
\bauthor{\bsnm{Paterson}, \binits{C.}},
\bauthor{\bsnm{Hawkins}, \binits{R.D.}},
\bauthor{\bsnm{Calinescu}, \binits{R.}},
\bauthor{\bsnm{Habli}, \binits{I.}}:
\bctitle{Assurance argument patterns and processes for machine learning in safety-related systems}.
In: \bbtitle{Proceedings of the Workshop on Artificial Intelligence Safety (SafeAI 2020)},
pp. \bfpage{23}--\blpage{30}
(\byear{2020}).
\bcomment{CEUR Workshop Proceedings}
\end{bchapter}
\endbibitem

\bibitem[\protect\citeauthoryear{Leveson}{2016}]{leveson2016engineering}
\begin{bbook}
\bauthor{\bsnm{Leveson}, \binits{N.G.}}:
\bbtitle{Engineering a Safer World: Systems Thinking Applied to Safety}.
\bpublisher{The MIT Press},
\blocation{Cambridge, MA, USA}
(\byear{2016})
\end{bbook}
\endbibitem

\bibitem[\protect\citeauthoryear{Borg et~al.}{2021}]{borg2021exploring}
\begin{bchapter}
\bauthor{\bsnm{Borg}, \binits{M.}},
\bauthor{\bsnm{Bronson}, \binits{J.}},
\bauthor{\bsnm{Christensson}, \binits{L.}},
\bauthor{\bsnm{Olsson}, \binits{F.}},
\bauthor{\bsnm{Lennartsson}, \binits{O.}},
\bauthor{\bsnm{Sonnsj{\"o}}, \binits{E.}},
\bauthor{\bsnm{Ebabi}, \binits{H.}},
\bauthor{\bsnm{Karsberg}, \binits{M.}}:
\bctitle{Exploring the assessment list for trustworthy ai in the context of advanced driver-assistance systems}.
In: \bbtitle{2021 IEEE/ACM 2nd International Workshop on Ethics in Software Engineering Research and Practice (SEthics)},
pp. \bfpage{5}--\blpage{12}
(\byear{2021}).
\bcomment{IEEE}
\end{bchapter}
\endbibitem

\bibitem[\protect\citeauthoryear{Gardner et~al.}{2022}]{gardner2022ethical}
\begin{botherref}
\oauthor{\bsnm{Gardner}, \binits{A.}},
\oauthor{\bsnm{Smith}, \binits{A.L.}},
\oauthor{\bsnm{Steventon}, \binits{A.}},
\oauthor{\bsnm{Coughlan}, \binits{E.}},
\oauthor{\bsnm{Oldfield}, \binits{M.}}:
Ethical funding for trustworthy ai: proposals to address the responsibilities of funders to ensure that projects adhere to trustworthy ai practice.
AI and Ethics,
1--15
(2022)
\end{botherref}
\endbibitem

\bibitem[\protect\citeauthoryear{Radclyffe et~al.}{2023}]{radclyffe2023assessment}
\begin{barticle}
\bauthor{\bsnm{Radclyffe}, \binits{C.}},
\bauthor{\bsnm{Ribeiro}, \binits{M.}},
\bauthor{\bsnm{Wortham}, \binits{R.H.}}:
\batitle{The assessment list for trustworthy artificial intelligence: A review and recommendations}.
\bjtitle{Frontiers in Artificial Intelligence}
\bvolume{6},
\bfpage{1020592}
(\byear{2023})
\end{barticle}
\endbibitem

\bibitem[\protect\citeauthoryear{Mitchell et~al.}{2019}]{mitchell2019model}
\begin{bchapter}
\bauthor{\bsnm{Mitchell}, \binits{M.}},
\bauthor{\bsnm{Wu}, \binits{S.}},
\bauthor{\bsnm{Zaldivar}, \binits{A.}},
\bauthor{\bsnm{Barnes}, \binits{P.}},
\bauthor{\bsnm{Vasserman}, \binits{L.}},
\bauthor{\bsnm{Hutchinson}, \binits{B.}},
\bauthor{\bsnm{Spitzer}, \binits{E.}},
\bauthor{\bsnm{Raji}, \binits{I.D.}},
\bauthor{\bsnm{Gebru}, \binits{T.}}:
\bctitle{Model cards for model reporting}.
In: \bbtitle{Proceedings of the Conference on Fairness, Accountability, and Transparency},
pp. \bfpage{220}--\blpage{229}
(\byear{2019})
\end{bchapter}
\endbibitem

\bibitem[\protect\citeauthoryear{Ashmore et~al.}{2021}]{ashmore2021assuring}
\begin{barticle}
\bauthor{\bsnm{Ashmore}, \binits{R.}},
\bauthor{\bsnm{Calinescu}, \binits{R.}},
\bauthor{\bsnm{Paterson}, \binits{C.}}:
\batitle{Assuring the machine learning lifecycle: Desiderata, methods, and challenges}.
\bjtitle{ACM Computing Surveys (CSUR)}
\bvolume{54}(\bissue{5}),
\bfpage{1}--\blpage{39}
(\byear{2021})
\end{barticle}
\endbibitem

\bibitem[\protect\citeauthoryear{{IEEE Computer Society/Artificial Intelligence Standards Committee (C/AISC)}}{2019}]{ieee2019draft}
\begin{bbook}
\bauthor{\bsnm{{IEEE Computer Society/Artificial Intelligence Standards Committee (C/AISC)}}}:
\bbtitle{IEEE Draft Standard for Responsible AI Licensing}
vol. \bseriesno{IEEE P2840}.
\bpublisher{{IEEE}},
\blocation{New York, USA}
(\byear{2019})
\end{bbook}
\endbibitem

\bibitem[\protect\citeauthoryear{{IEEE Computer Society/Artificial Intelligence Standards Committee (C/AISC)}}{2020}]{ieee2020recommended}
\begin{bbook}
\bauthor{\bsnm{{IEEE Computer Society/Artificial Intelligence Standards Committee (C/AISC)}}}:
\bbtitle{Recommended Practice for Organizational Governance of Artificial Intelligence}
vol. \bseriesno{IEEE P2863}.
\bpublisher{{IEEE}},
\blocation{New York, USA}
(\byear{2020})
\end{bbook}
\endbibitem

\bibitem[\protect\citeauthoryear{Schwartz et~al.}{2022}]{schwartz2022towards}
\begin{bbook}
\bauthor{\bsnm{Schwartz}, \binits{R.}},
\bauthor{\bsnm{Schwartz}, \binits{R.}},
\bauthor{\bsnm{Vassilev}, \binits{A.}},
\bauthor{\bsnm{Greene}, \binits{K.}},
\bauthor{\bsnm{Perine}, \binits{L.}},
\bauthor{\bsnm{Burt}, \binits{A.}},
\bauthor{\bsnm{Hall}, \binits{P.}}:
\bbtitle{Towards a Standard for Identifying and Managing Bias in Artificial Intelligence}
vol. \bseriesno{3}.
\bpublisher{US Department of Commerce, National Institute of Standards and Technology},
\blocation{USA}
(\byear{2022})
\end{bbook}
\endbibitem

\bibitem[\protect\citeauthoryear{{National Artificial Intelligence Centre}}{2024}]{national2024voluntary}
\begin{bbook}
\bauthor{\bsnm{{National Artificial Intelligence Centre}}}:
\bbtitle{Voluntary AI Safety Standard --- Guiding Safe and Responsible Use of Artificial Intelligence in Australia}.
\bpublisher{{Australian Department of Industry, Science and Resources}},
\blocation{Canberra, Australia}
(\byear{2024})
\end{bbook}
\endbibitem

\bibitem[\protect\citeauthoryear{Liao et~al.}{2021}]{liao2021question}
\begin{botherref}
\oauthor{\bsnm{Liao}, \binits{Q.V.}},
\oauthor{\bsnm{Pribi{\'c}}, \binits{M.}},
\oauthor{\bsnm{Han}, \binits{J.}},
\oauthor{\bsnm{Miller}, \binits{S.}},
\oauthor{\bsnm{Sow}, \binits{D.}}:
Question-driven design process for explainable ai user experiences.
arXiv preprint arXiv:2104.03483
(2021)
\end{botherref}
\endbibitem

\bibitem[\protect\citeauthoryear{Duan et~al.}{2017}]{duan2017reasoning}
\begin{bchapter}
\bauthor{\bsnm{Duan}, \binits{L.}},
\bauthor{\bsnm{Rayadurgam}, \binits{S.}},
\bauthor{\bsnm{Heimdahl}, \binits{M.P.}},
\bauthor{\bsnm{Ayoub}, \binits{A.}},
\bauthor{\bsnm{Sokolsky}, \binits{O.}},
\bauthor{\bsnm{Lee}, \binits{I.}}:
\bctitle{Reasoning about confidence and uncertainty in assurance cases: A survey}.
In: \bbtitle{Software Engineering in Health Care: 4th International Symposium, FHIES 2014, and 6th International Workshop, SEHC 2014, Washington, DC, USA, July 17-18, 2014, Revised Selected Papers 4},
pp. \bfpage{64}--\blpage{80}
(\byear{2017}).
\bcomment{Springer}
\end{bchapter}
\endbibitem

\bibitem[\protect\citeauthoryear{Grigorova and Maibaum}{2013}]{grigorova2013taking}
\begin{bchapter}
\bauthor{\bsnm{Grigorova}, \binits{S.}},
\bauthor{\bsnm{Maibaum}, \binits{T.}}:
\bctitle{Taking a page from the law books: Considering evidence weight in evaluating assurance case confidence}.
In: \bbtitle{2013 IEEE International Symposium on Software Reliability Engineering Workshops (ISSREW)},
pp. \bfpage{387}--\blpage{390}
(\byear{2013}).
\bcomment{IEEE}
\end{bchapter}
\endbibitem

\bibitem[\protect\citeauthoryear{Maksimov et~al.}{2019}]{maksimov2019survey}
\begin{barticle}
\bauthor{\bsnm{Maksimov}, \binits{M.}},
\bauthor{\bsnm{Kokaly}, \binits{S.}},
\bauthor{\bsnm{Chechik}, \binits{M.}}:
\batitle{A survey of tool-supported assurance case assessment techniques}.
\bjtitle{ACM Computing Surveys (CSUR)}
\bvolume{52}(\bissue{5}),
\bfpage{1}--\blpage{34}
(\byear{2019})
\end{barticle}
\endbibitem

\bibitem[\protect\citeauthoryear{Mohamad et~al.}{2021}]{mohamad2021security}
\begin{barticle}
\bauthor{\bsnm{Mohamad}, \binits{M.}},
\bauthor{\bsnm{Stegh{\"o}fer}, \binits{J.-P.}},
\bauthor{\bsnm{Scandariato}, \binits{R.}}:
\batitle{Security assurance cases—state of the art of an emerging approach}.
\bjtitle{Empirical software engineering}
\bvolume{26}(\bissue{4}),
\bfpage{70}
(\byear{2021})
\end{barticle}
\endbibitem

\bibitem[\protect\citeauthoryear{Shahandashti et~al.}{2023}]{shahandashti2023prisma}
\begin{botherref}
\oauthor{\bsnm{Shahandashti}, \binits{K.K.}},
\oauthor{\bsnm{Belle}, \binits{A.B.}},
\oauthor{\bsnm{Lethbridge}, \binits{T.C.}},
\oauthor{\bsnm{Odu}, \binits{O.}},
\oauthor{\bsnm{Sivakumar}, \binits{M.}}:
A prisma-driven systematic mapping study on system assurance weakeners.
arXiv preprint arXiv:2311.08328
(2023)
\end{botherref}
\endbibitem

\bibitem[\protect\citeauthoryear{C{\^a}rlan et~al.}{2022}]{carlan2022automating}
\begin{bchapter}
\bauthor{\bsnm{C{\^a}rlan}, \binits{C.}},
\bauthor{\bsnm{Gauerhof}, \binits{L.}},
\bauthor{\bsnm{Gallina}, \binits{B.}},
\bauthor{\bsnm{Burton}, \binits{S.}}:
\bctitle{Automating safety argument change impact analysis for machine learning components}.
In: \bbtitle{2022 IEEE 27th Pacific Rim International Symposium on Dependable Computing (PRDC)},
pp. \bfpage{43}--\blpage{53}
(\byear{2022}).
\bcomment{IEEE}
\end{bchapter}
\endbibitem

\bibitem[\protect\citeauthoryear{Yap}{2021}]{yap2021towards}
\begin{bchapter}
\bauthor{\bsnm{Yap}, \binits{R.H.}}:
\bctitle{Towards certifying trustworthy machine learning systems}.
In: \bbtitle{Trustworthy AI-Integrating Learning, Optimization and Reasoning: First International Workshop, TAILOR 2020, Virtual Event, September 4--5, 2020, Revised Selected Papers 1},
pp. \bfpage{77}--\blpage{82}
(\byear{2021}).
\bcomment{Springer}
\end{bchapter}
\endbibitem

\bibitem[\protect\citeauthoryear{15026-2-2022}{}]{ieee2022standard}
\begin{botherref}
\oauthor{\bsnm{15026-2-2022}, \binits{I.}}:
ISO/IEC/IEEE International Standard - Systems and software engineering--Systems and software assurance--Part 2: Assurance case.
https://standards.ieee.org/ieee/15026-2/10236/
\end{botherref}
\endbibitem

\bibitem[\protect\citeauthoryear{Mansourov and Campara}{2010}]{mansourov2010system}
\begin{bbook}
\bauthor{\bsnm{Mansourov}, \binits{N.}},
\bauthor{\bsnm{Campara}, \binits{D.}}:
\bbtitle{System Assurance: Beyond Detecting Vulnerabilities}.
\bpublisher{Morgan Kaufmann},
\blocation{MA, USA}
(\byear{2010})
\end{bbook}
\endbibitem

\bibitem[\protect\citeauthoryear{Chelouati et~al.}{2023}]{chelouati2023graphical}
\begin{barticle}
\bauthor{\bsnm{Chelouati}, \binits{M.}},
\bauthor{\bsnm{Boussif}, \binits{A.}},
\bauthor{\bsnm{Beugin}, \binits{J.}},
\bauthor{\bsnm{El~Koursi}, \binits{E.-M.}}:
\batitle{Graphical safety assurance case using goal structuring notation (gsn)—challenges, opportunities and a framework for autonomous trains}.
\bjtitle{Reliability Engineering \& System Safety}
\bvolume{230},
\bfpage{108933}
(\byear{2023})
\end{barticle}
\endbibitem

\bibitem[\protect\citeauthoryear{Horkoff et~al.}{2019}]{horkoff2019goal}
\begin{barticle}
\bauthor{\bsnm{Horkoff}, \binits{J.}},
\bauthor{\bsnm{Aydemir}, \binits{F.B.}},
\bauthor{\bsnm{Cardoso}, \binits{E.}},
\bauthor{\bsnm{Li}, \binits{T.}},
\bauthor{\bsnm{Mat{\'e}}, \binits{A.}},
\bauthor{\bsnm{Paja}, \binits{E.}},
\bauthor{\bsnm{Salnitri}, \binits{M.}},
\bauthor{\bsnm{Piras}, \binits{L.}},
\bauthor{\bsnm{Mylopoulos}, \binits{J.}},
\bauthor{\bsnm{Giorgini}, \binits{P.}}:
\batitle{Goal-oriented requirements engineering: an extended systematic mapping study}.
\bjtitle{Requirements engineering}
\bvolume{24},
\bfpage{133}--\blpage{160}
(\byear{2019})
\end{barticle}
\endbibitem

\bibitem[\protect\citeauthoryear{ITU-T~Z.151}{2018}]{itu2018151}
\begin{botherref}
\oauthor{\bsnm{ITU-T~Z.151}, \binits{I.-T.R.}}:
Z.151 user requirements notation (urn)--language definition.
ITUT, Oct
(2018)
\end{botherref}
\endbibitem

\bibitem[\protect\citeauthoryear{Dongmo and Van~der Poll}{2023}]{dongmo2023improved}
\begin{barticle}
\bauthor{\bsnm{Dongmo}, \binits{C.}},
\bauthor{\bsnm{Poll}, \binits{J.A.}}:
\batitle{An improved user requirements notation (urn) models’ construction approach}.
\bjtitle{Systems}
\bvolume{11}(\bissue{6}),
\bfpage{301}
(\byear{2023})
\end{barticle}
\endbibitem

\bibitem[\protect\citeauthoryear{Amyot and Mussbacher}{2011}]{amyot2011user}
\begin{barticle}
\bauthor{\bsnm{Amyot}, \binits{D.}},
\bauthor{\bsnm{Mussbacher}, \binits{G.}}:
\batitle{User requirements notation: the first ten years, the next ten years}.
\bjtitle{Journal of Software}
\bvolume{6}(\bissue{5}),
\bfpage{747}--\blpage{768}
(\byear{2011})
\end{barticle}
\endbibitem

\bibitem[\protect\citeauthoryear{Weiss and Amyot}{2007}]{weiss2007business}
\begin{bchapter}
\bauthor{\bsnm{Weiss}, \binits{M.}},
\bauthor{\bsnm{Amyot}, \binits{D.}}:
\bctitle{Business model design and evolution}.
In: \bbtitle{Challenges in the Management of New Technologies},
pp. \bfpage{183}--\blpage{194}.
\bpublisher{World Scientific},
\blocation{Singapore}
(\byear{2007})
\end{bchapter}
\endbibitem

\bibitem[\protect\citeauthoryear{Canca}{}]{canca2019new}
\begin{botherref}
\oauthor{\bsnm{Canca}, \binits{C.}}:
A {New} {Model} {For} {AI} {Ethics} {In} {R}\&{D}.
\url{https://www.forbes.com/sites/insights-intelai/2019/03/27/rethinking-ethics-in-ai-rd/}
Accessed 2019-10-23
\end{botherref}
\endbibitem

\bibitem[\protect\citeauthoryear{Zhou et~al.}{2021}]{zhou2021evaluating}
\begin{barticle}
\bauthor{\bsnm{Zhou}, \binits{J.}},
\bauthor{\bsnm{Gandomi}, \binits{A.H.}},
\bauthor{\bsnm{Chen}, \binits{F.}},
\bauthor{\bsnm{Holzinger}, \binits{A.}}:
\batitle{Evaluating the quality of machine learning explanations: A survey on methods and metrics}.
\bjtitle{Electronics}
\bvolume{10}(\bissue{5}),
\bfpage{593}
(\byear{2021})
\end{barticle}
\endbibitem

\bibitem[\protect\citeauthoryear{Bellamy et~al.}{2018}]{Bellamy_fairness_2018}
\begin{botherref}
\oauthor{\bsnm{Bellamy}, \binits{R.K.E.}},
\oauthor{\bsnm{Dey}, \binits{K.}},
\oauthor{\bsnm{Hind}, \binits{M.}},
\oauthor{\bsnm{Hoffman}, \binits{S.C.}},
\oauthor{\bsnm{Houde}, \binits{S.}},
\oauthor{\bsnm{Kannan}, \binits{K.}},
\oauthor{\bsnm{Lohia}, \binits{P.}},
\oauthor{\bsnm{Martino}, \binits{J.}},
\oauthor{\bsnm{Mehta}, \binits{S.}},
\oauthor{\bsnm{Mojsilovic}, \binits{A.}},
\oauthor{\bsnm{Nagar}, \binits{S.}},
\oauthor{\bsnm{Ramamurthy}, \binits{K.N.}},
\oauthor{\bsnm{Richards}, \binits{J.T.}},
\oauthor{\bsnm{Saha}, \binits{D.}},
\oauthor{\bsnm{Sattigeri}, \binits{P.}},
\oauthor{\bsnm{Singh}, \binits{M.}},
\oauthor{\bsnm{Varshney}, \binits{K.R.}},
\oauthor{\bsnm{Zhang}, \binits{Y.}}:
{AI} fairness 360: An extensible toolkit for detecting, understanding, and mitigating unwanted algorithmic bias.
CoRR
\textbf{abs/1810.01943}
(2018)
{\href{https://arxiv.org/abs/1810.01943}{{arXiv:1810.01943}}}
\end{botherref}
\endbibitem

\bibitem[\protect\citeauthoryear{Kairouz et~al.}{2021}]{kairouz2021advances}
\begin{barticle}
\bauthor{\bsnm{Kairouz}, \binits{P.}},
\bauthor{\bsnm{McMahan}, \binits{H.B.}},
\bauthor{\bsnm{Avent}, \binits{B.}},
\bauthor{\bsnm{Bellet}, \binits{A.}},
\bauthor{\bsnm{Bennis}, \binits{M.}},
\bauthor{\bsnm{Bhagoji}, \binits{A.N.}},
\bauthor{\bsnm{Bonawitz}, \binits{K.}},
\bauthor{\bsnm{Charles}, \binits{Z.}},
\bauthor{\bsnm{Cormode}, \binits{G.}},
\bauthor{\bsnm{Cummings}, \binits{R.}}, \betal:
\batitle{Advances and open problems in federated learning}.
\bjtitle{Foundations and Trends{\textregistered} in Machine Learning}
\bvolume{14}(\bissue{1--2}),
\bfpage{1}--\blpage{210}
(\byear{2021})
\end{barticle}
\endbibitem

\bibitem[\protect\citeauthoryear{Morley et~al.}{2020}]{morley2020initial}
\begin{barticle}
\bauthor{\bsnm{Morley}, \binits{J.}},
\bauthor{\bsnm{Floridi}, \binits{L.}},
\bauthor{\bsnm{Kinsey}, \binits{L.}},
\bauthor{\bsnm{Elhalal}, \binits{A.}}:
\batitle{From what to how: an initial review of publicly available ai ethics tools, methods and research to translate principles into practices}.
\bjtitle{Science and engineering ethics}
\bvolume{26}(\bissue{4}),
\bfpage{2141}--\blpage{2168}
(\byear{2020})
\end{barticle}
\endbibitem

\bibitem[\protect\citeauthoryear{Gusdorf}{2008}]{gusdorf2008recruitment}
\begin{botherref}
\oauthor{\bsnm{Gusdorf}, \binits{M.L.}}:
Recruitment and selection: Hiring the right person.
USA: Society for Human Resource Management,
1--14
(2008)
\end{botherref}
\endbibitem

\bibitem[\protect\citeauthoryear{Kaushal et~al.}{2023}]{kaushal2023artificial}
\begin{barticle}
\bauthor{\bsnm{Kaushal}, \binits{N.}},
\bauthor{\bsnm{Kaurav}, \binits{R.P.S.}},
\bauthor{\bsnm{Sivathanu}, \binits{B.}},
\bauthor{\bsnm{Kaushik}, \binits{N.}}:
\batitle{Artificial intelligence and hrm: identifying future research agenda using systematic literature review and bibliometric analysis}.
\bjtitle{Management Review Quarterly}
\bvolume{73}(\bissue{2}),
\bfpage{455}--\blpage{493}
(\byear{2023})
\end{barticle}
\endbibitem

\bibitem[\protect\citeauthoryear{Roopalatha and Sucharita}{2024}]{roopalatha2024artificial}
\begin{barticle}
\bauthor{\bsnm{Roopalatha}, \binits{N.}},
\bauthor{\bsnm{Sucharita}, \binits{K.}}:
\batitle{Artificial intelligence on human resource management-innovation, challenges and path forward}.
\bjtitle{Educational Administration: Theory and Practice}
\bvolume{30}(\bissue{5}),
\bfpage{13686}--\blpage{13698}
(\byear{2024})
\end{barticle}
\endbibitem

\bibitem[\protect\citeauthoryear{Ajonbadi et~al.}{2024}]{ajonbadi2024before}
\begin{bchapter}
\bauthor{\bsnm{Ajonbadi}, \binits{H.A.}},
\bauthor{\bsnm{Khan}, \binits{S.}},
\bauthor{\bsnm{Owolewa}, \binits{M.}}:
\bctitle{Before it goes south: The ethical dilemma of artificial intelligence in human resource management—the bangladesh workplace experience}.
In: \bbtitle{HRM, Artificial Intelligence and the Future of Work: Insights from the Global South},
pp. \bfpage{147}--\blpage{167}.
\bpublisher{Palgrave Macmillan},
\blocation{Cham, Switzerland}
(\byear{2024})
\end{bchapter}
\endbibitem

\end{thebibliography}

\end{document}